\documentclass[final,3p,times]{elsarticle}

\usepackage{mathrsfs}
\usepackage{amssymb}
\usepackage{amsmath}
\usepackage{amsthm}
\usepackage{physics}

\usepackage{lipsum}
\usepackage{graphicx}
\usepackage{dcolumn}
\usepackage{bm}
\usepackage{multirow}
\usepackage{comment}
\usepackage{subfigure}
\usepackage{soul}

\usepackage[usenames,dvipsnames]{color}
\usepackage{hyperref}

\usepackage[mathlines]{lineno}

\usepackage{setspace}

\usepackage{verbatim} 

\DeclareGraphicsExtensions{.pdf,.png,.jpg}

\usepackage{natbib}
\journal{Nuclear~Instruments~and~Methods~A}

\usepackage{setspace,caption}
\usepackage{lipsum}

\begin{document}

\newcommand{\bmd}{$\beta$}
\newcommand{\bpd}{$\beta^{+}$}
\newcommand{\bb}{$\beta\beta$}
\newcommand{\nbb}{\ensuremath{\nu\beta\beta}}
\newcommand{\vbb}{\ensuremath{0\nu\beta\beta}}
\newcommand{\Te}{$^{130}$Te}
\newcommand{\Se}{$^{82}$Se}
\newcommand{\B}{$^{8}$B }
\newcommand{\Bten}{$^{10}$B}
\newcommand{\Cten}{$^{10}$C}



\title{Space-Time Discriminant to Separate Double-Beta Decay from $^8$B Solar Neutrinos in Liquid Scintillator}


\author{Runyu Jiang and Andrey Elagin\corref{cor1}}
\cortext[cor1]{Corresponding Author: \href{mailto:elagin@hpe.uchicago.edu}{\tt{elagin@hep.uchicago.edu}}}

\address{Enrico Fermi Institute, University of Chicago, Chicago, IL 60637 }

\setstretch{1.0}

\begin{abstract}

We present a technique for separating nuclear double beta-decay events
from background \B solar neutrino interactions in a liquid
scintillator detector instrumented with photo-detectors with mm space
and 100-psec time resolutions. The technique uses position and time
information of detected photo-electrons (PEs) to separate directional
Cherenkov light from isotropic scintillation light in the
reconstruction of the kinematics of candidate events. Here we
introduce a Cherenkov-scintillation space time boundary defined as the 
light cone in the 2-dimensional space of the arrival time and the polar angle of 
each PE with respect to the axis from the center of
the detector to the vertex. The PEs located near the boundary correspond to
photons that were emitted early and contain a high fraction of directional 
Cherenkov PEs. We apply weights derived from the distance to the
boundary of each individual PE, which are then used in a spherical
harmonics analysis that separates the two-track event topology of
double-beta decay from the one-track topology of $^8$B events.
The Geant-4 simulation assumes a detector of 6.5 m radius filled with $^{130}$Te-loaded 
liquid scintillator and surrounded by photo-detectors with time and space resolutions 
of 100 ps and 3~mm respectively. The scintillation properties and photo-detector quantum
efficiency are modeled after KamLAND. Assuming a fiducial volume of 3 m radius, 
a photo-coverage of 65\% and vertex resolution of $\sigma_{r_v} = $ 5.2 cm at 2.53 MeV
the method of reconstructing event topology 
predicts factors of 1.3 and 2.3 in background suppression at 90\% and 70\% 
signal efficiency respectively. Additionally, the PEs near the Cherenkov-scintillation space-time boundary can be used to reconstruct the directionality of one-electron candidate events, allowing for further $^8$B background suppression due to the correlation between the direction of the scattered electron and the position of the sun. We find polar and azimuthal
angular resolutions of 0.46 and 0.84 radians respectively. We show the dependence of the topology and directionality reconstructions on photo-coverage and
vertex resolution, and discuss directions in detector development that
can improve background suppression; however determination of a combined background rejection factor based on topological and directional reconstruction is a subject of further studies using a detailed detector-specific background model.

\end{abstract}
\setstretch{1.0}
\date{\today}

\maketitle
\clearpage
\tableofcontents

\setstretch{1.0}

\clearpage
\section{Introduction}

Today it is still not known whether for a given helicity the neutrino
mass eigenstates are identical to the corresponding eigenstates of the anti-neutrino, i.e. whether the neutrino is 
a Majorana particle~\cite{Majorana1937}.

Searching for neutrinoless double-beta decay (\vbb-decay) \cite{Furry1939} is the most 
feasible experimental technique to determine if the 
neutrino is a Majorana particle. Contrary to two-neutrino double-beta decay~\cite{GoeppertMayer1935},
neutrinoless double-beta decay, $(Z,A)\rightarrow(Z+2,A)+2e^-$, violates electron lepton number by two units and 
therefore has a potential to probe physics beyond the Standard Model~\cite{Duerr:2013iwa}. At the same time, according 
to the Schechter-Valle theorem~\cite{Schechter_Valle_theorem}, the observation of \vbb-decay would guarantee the 
existence of a non-zero Majorana mass term for the neutrino\footnote{For a ``natural'' gauge theory without an extreme 
fine tuning~\cite{Schechter_Valle_theorem}. See also Ref.~\cite{Duerr:2011zd} for more discussion on the 
Schechter-Valle theorem.}.

To explore \vbb-decay beyond the current experimental lower limits on half-life of up to $\sim$$10^{26}$ years~\cite{EXO2014, KLZ2016, GERDA2018, MAJORANA-2019}, all currently planned \vbb-decay experiments aim for a ton-scale active isotope 
mass~\cite{NSACreport}. There are several detector technologies capable of reaching a sensitivity to a \vbb-decay half-life of $10^{27}$-$10^{28}$\, years by instrumenting several tons of isotope \cite{nEXO_PRC2018, LEGEND_AIPConf2017, giuliani_andrea_2018_1286915}.  Probing Majorana masses in the regime of the non-degenerate normal neutrino mass hierarchy may require a
detector with an active isotope mass approaching a kilo-ton~\cite{Biller2013}.

The scalability, self-shielding, and good energy resolution of liquid scintillator detectors makes them a competitive option 
for the search for \vbb-decay~\cite{KLZ2016}. The use of Cherenkov light in a liquid scintillator detector can enhance the event identification
capabilities of liquid scintillator detectors. Cherenkov light has been successfully used for event reconstruction in a diluted liquid 
scintillator in the LSND experiment~\cite{LSND_diluted_scintillator}. However, the light yield of the diluted scintillator of LSND would not be
sufficient to achieve the energy resolution required for \vbb-decay searches. In addition, LSND was looking for electron tracks with energy
of about 45 MeV. This is to be compared with the $\sim$1-2~MeV electrons from a \vbb-decay. Reconstruction of events with energies down to 3-5 MeV using Cherenkov light in
pure water has been done at the Super-Kamiokande experiment~\cite{Smy_bibtex_from_inspire, Smy_Method_in_PRD}. In this work we focus on the reconstruction of \vbb-decay events in a high light-yield liquid scintillator.

To our knowledge, the idea of using Cherenkov light in searches for \vbb-decay was first discussed in Ref.~\cite{qdot}.
Development of a scintillator with a long time constant to extract directional information from Cherenkov light in \vbb-decay searches 
has been suggested in Ref.~\cite{Biller2013}.
The first feasibility studies of using Cherenkov light in a regular liquid scintillator with the scintillation light yield sufficient for
\vbb-decay search were done in Ref.~\cite{Aberle2014}. It was shown that by correlating PE position and time measurements using fast 
photo-detectors with time resolution of $\sim$100~ps one can reconstruct the direction of electrons in the energy range between
1.4 and 5~MeV. This in turn, opens a possibility to suppress backgrounds in \vbb-decay searches, including background from 
\B solar neutrinos.

In a kilo-ton liquid scintillator detector, the electrons from elastic scattering 
of \B solar neutrinos can become a dominant background~\cite{snoplus2016}.
A path towards suppression of \B background has been shown in previous work~\cite{Aberle2014, harmonics2017}. The background suppression relies on separation of directional Cherenkov light from the abundant isotropic scintillation light using fast photo-detectors. Cherenkov photons then allow reconstruction of the event topology, which is different for the two-electron \vbb-decay signal and for one-electron \B background events~\cite{harmonics2017}. Whenever the \vbb-decay signal events have only one electron above Cherenkov threshold or have two electrons emitted at a very small angle, the \B background can be suppressed by reconstructing the direction of the electron(s), which in the case of \B events correlates with the position of the sun~\cite{Biller2013,theia_white_paper}.

Separation of Cherenkov light from scintillation light has been demonstrated in various experimental settings~\cite{LSND_diluted_scintillator, chessPRC, chessEPJC, Minfang_slow_rise_time} including detection of Cherenkov light from 1-2 MeV electrons in linear alkylbenzene~\cite{ring_paper2018}. Currently, research towards an effective separation of Cherenkov light from scintillation light is being actively pursued~\cite{Guo:2017, Wang:2017, Zhi:2017, Gooding:2018, Cumming:2018, UPenn_spectral_paper, ZICOS-2015, ZICOS-2018}. For example, the ANNIE experiment~\cite{ANNIE_physics_phase_proposal} plans to fill their detector volume with a water-based liquid scintillator~\cite{Minfang_WbLS_1, Minfang_WbLS_2} to explore a hybrid event reconstruction scheme that uses Cherenkov and scintillation light for events with energy around 1 GeV~\cite{ANNIE_private}. The NuDot experiment aims to demonstrate kinematics reconstruction in a one-ton liquid scintillator detector for events in the energy range relevant for \vbb-decay~\cite{ring_paper2018}.

New techniques have been developed for event reconstruction in liquid scintillator detectors to complement calorimetric measurements~\cite{Sakai_thesis, Gratta_imaging2018, Wonsak2018} including those targeting separation of \vbb-decay from cosmic muon spallation background~\cite{ML_C10_paper2018}.

Here we present new developments in the reconstruction of the \vbb-decay event topology and in measuring electron directionality for \B background suppression. Continuing the work described in Ref.~\cite{harmonics2017} we use spherical functions to construct topology-dependent rotation invariants. For directionality reconstruction, similarly to Ref.~\cite{Aberle2014}, we also use the ``center of gravity'' of a photo-electron (PE) sample with an enhanced fraction of Cherenkov light. However, the technique presented here is an improvement over Refs.~\cite{Aberle2014, harmonics2017} in the following areas.

First, instead of applying a fixed time cut in the case of central events~\cite{Aberle2014} or a pre-defined {\it ad-hoc} time window in the case of events uniformly distributed throughout the entire fiducial volume~\cite{harmonics2017} to separate Cherenkov and scintillation light, we identify a Cherenkov-scintillation space-time boundary with a general expression as the light cone in the 2-dimensional plane of the arrival time and the polar angle with respect to the axis from the center of the detector to the vertex, which allows selecting a PE sample with an increased fraction of Cherekov PEs.
In this paper, 
the time displacement of each PE 
from the Cherenkov-scintillation space time boundary is used to assign higher weights to Cherenkov PEs.

Second, we construct the spherical harmonics power spectrum, called here the S-spectrum\footnote{See Section~\ref{subsec:S-spectrum} and~\ref{app:harmonics} for the definition of the S-spectrum.}, using a summation over each individual PE instead of 
integrating over $\Delta \theta \times \Delta \phi$ segments, which in Ref.\cite{harmonics2017} on average have a solid angle of $\sim$0.063 corresponding to the surface area of 2.65~m$^2$ for a 6.5m radius sphere. This allows for the efficient use of spatial information of photo-electrons. At the same time this improves the CPU time needed to process one event compared to the technique in Ref.\cite{harmonics2017}, because there is no numerical integration involved in the calculation of the S-spectrum in the new method.

Third, while the reconstruction of electron directionality has been demonstrated in Ref.~\cite{Aberle2014} for the case of central events, here we demonstrate electron directionality reconstruction for all events uniformly distributed throughout the fiducial volume of the detector.

Finally, we use a maximum likelihood method to quantify the background suppression that can be obtained from two different variables derived from the S-spectrum. This is a step towards a multi-variative analysis to extract more information from the S-spectrum and to account for additional correlations between the S-spectrum components and other attributes of \vbb-decay candidate
events such as the position of the vertex, and the angle and the energy split between the electrons.

While we targeted the \B solar neutrino background suppression, the topological reconstruction is also applicable for the suppression of backgrounds due to one-track single-beta decays. The directionality reconstruction can be instrumental for studying CNO-cycle solar neutrinos~\cite{Bonventre:2018hyd} and geo-neutrinos~\cite{SDye_ncomms2017}.

The organization of the paper is as follows. Section~\ref{sec:detector_simulation} describes
the detector model and event simulation details. The Cherenkov-scintillation space-time boundary is introduced in Section~\ref{sec:boundary}. Section~\ref{sec:topology} describes \B background suppression based on event topology reconstruction. Section~\ref{sec:directionality} describes directionality reconstruction. 
Conclusions are summarized in Section~\ref{sec:conclusions}.

\section{Detector Model and Event Simulation}
\label{sec:detector_simulation}

We use the same Geant-4 based simulation described in Refs.~\cite{Aberle2014,harmonics2017}. The detector is a sphere with a radius 
of 6.5~m filled with liquid scintillator. The scintillator composition has been chosen to match a KamLAND-like 
scintillator~\cite{kamland2003}, which consists of 80\% n-dodecane,
20\% pseudocumene and 1.52~g/l PPO with a density of $\rho$ =
0.78~g/ml. However, we deviate from the baseline KamLAND case in that the re-emission of absorbed photons 
in the scintillator bulk volume and optical scattering, specifically Rayleigh scattering, have not yet been included. A test 
simulation shows that the effect of optical scattering is negligible~\cite{Aberle2014}.

The inner sphere surface is used as the photo-detector. It is treated as fully absorbing with no reflections. We consider 
photo-cathode coverage ranging from 10\% to 100\% with 65\% being our default coverage choice. We distribute 7682 identical spherical caps over the detector sphere and insert inside identical circular photo-detectors. We vary the radii of the photo-detectors from 47~mm to 120~mm to achieve a desired photo-coverage\footnote{See~\ref{app:coverage} for a detailed description of the photo-coverage implementation}.

The assumed quantum efficiency (QE) is that of a typical bialkali photo-cathode 
(Hamamatsu R7081 PMT~\cite{Hamamatsu_R7081}, see also Ref.~\cite{dctwo}), which is 12\% for Cherenkov light and 23\% for 
scintillation light. We assume a photo-detector transit-time-spread (TTS) of 100~ps, and PE position resolution of 3~mm in both of the two dimensions, which,
for example, are well within demonstrated performance of large-area picosecond photo-detectors (LAPPD)~\cite{Timing_paper, Incom_paper}.
We neglect the threshold effects in the photo-detector readout electronics, noise effects, and contributions to time resolution other 
than the photo-detector TTS.

We define the fiducial volume as a sphere of 3~m radius and simulate signal and background events uniformly distributed within that fiducial volume. Following the same strategy as in Ref.~\cite{harmonics2017} we smear the vertex along $x$, $y$, and $z$ directions with three independent Gaussian distributions of the same width, $\sigma_x = \sigma_y = \sigma_z =$ 3~cm.
This vertex smearing introduces an uncertainty on the vertex position of $\sigma_{r_v} = $ 5.2~cm, which is based on an earlier study of vertex reconstruction~\cite{Aberle2014}. The vertex position,
$r_v$, is then treated as a reconstructed vertex position.

The details of the event simulation and discussion on kinematics of $^{130}$Te \vbb-decay and \B events can be found in Ref.~\cite{harmonics2017}.
We simulate the kinematics of 0\nbb-decay events using a custom Monte Carlo with momentum and angle-dependent phase space
factors for 0\nbb-decay~\cite{Jenni}.
Electrons from elastic scattering of \B solar neutrinos have a nearly flat energy spectrum around the \Te~\vbb-decay Q-value of 2.53~MeV\cite{SNOp-B8-bkg}. 
We simulate \B background as a single monochromatic electron with energy of 2.53~MeV.

\section{Cherenkov-Scintillation Space-Time Boundary}
\label{sec:boundary}

For any given sufficiently small area on the detector surface that is hit by the Cherenkov light, the majority of the scintillation PEs arrive after the Cherenkov PEs regardless of the vertex position and the event type. In the 2-dimensional plane of the arrival time and the polar angle with respect to the axis from the center of the detector to the vertex, the Cherenkov-scintillation space-time boundary $t_{c}(\theta)$ is defined to be the light cone with a certain speed of light:

\begin{eqnarray}
\label{eq:def_time_boundary}
	t_{c}(\theta)=\frac{\sqrt{R^2+r_v^2-2Rr_v \cos\theta}}{c(n)}-\frac{R-r_v}{c(n)},
\end{eqnarray}
where $R$ is the radius of the detector's inner sphere, $r_v$ is the length of the vertex displacement vector $\vec{r}_v$ from the center of the sphere, $\theta$ is the angle between the vertex displacement vector $\vec{r}_v$ and the photon hit vector from the center of the sphere, and $c(n)$ is the speed of the scintillation light in the liquid scintillator. We use an average value for the index of refraction of $n = $ 1.53 to determine $c(n)$. The majority of scintillation PEs have arrival times, $t_{sci}(\theta)$, exceeding $t_c(\theta)$, $t_{sci}(\theta) > t_c(\theta)$. 

The first term in Eq.\ref{eq:def_time_boundary} corresponds to an estimate of the time for the first scintillation PE emitted at angle $\theta$ to reach the detector sphere. The second term is an estimate of the travel time for the earliest scintillation PEs, i.e. the scintillation PE emitted at $\theta = 0$, along the vertex displacement vector $\vec{r}_v$.

\begin{figure}[h!]
    \includegraphics[width=0.49\textwidth]{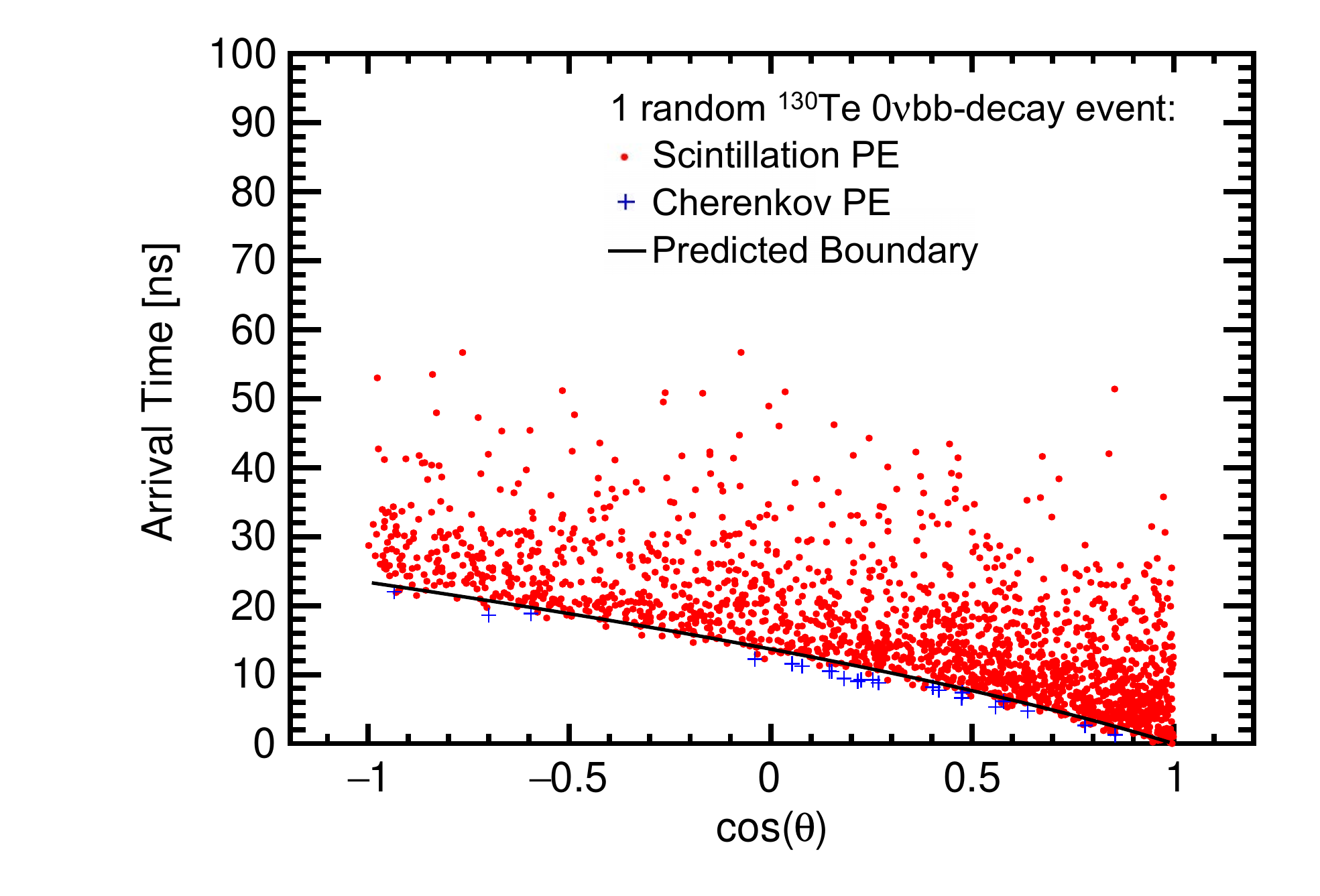}
    \includegraphics[width=0.49\textwidth]{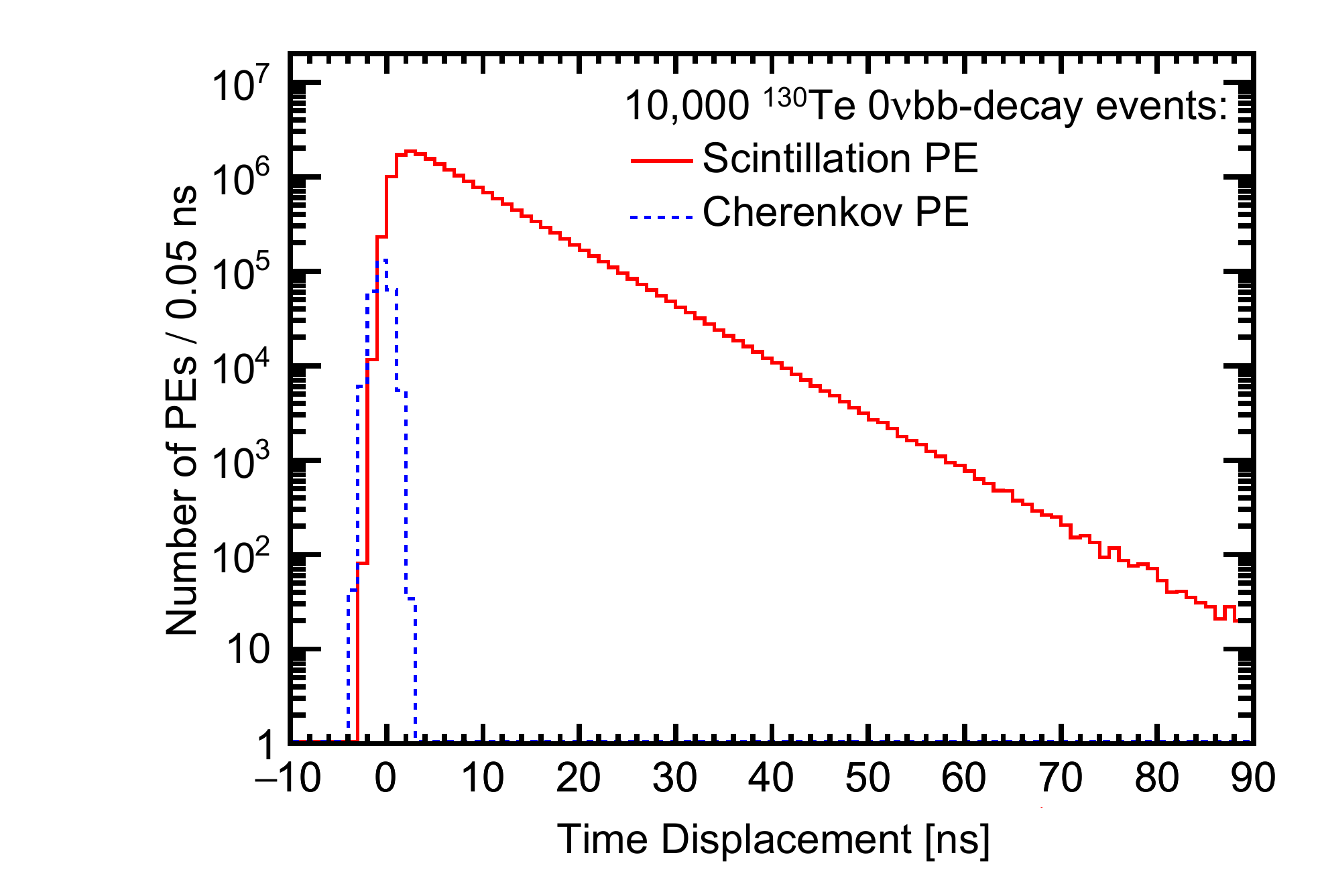}
    \includegraphics[width=0.49\textwidth]{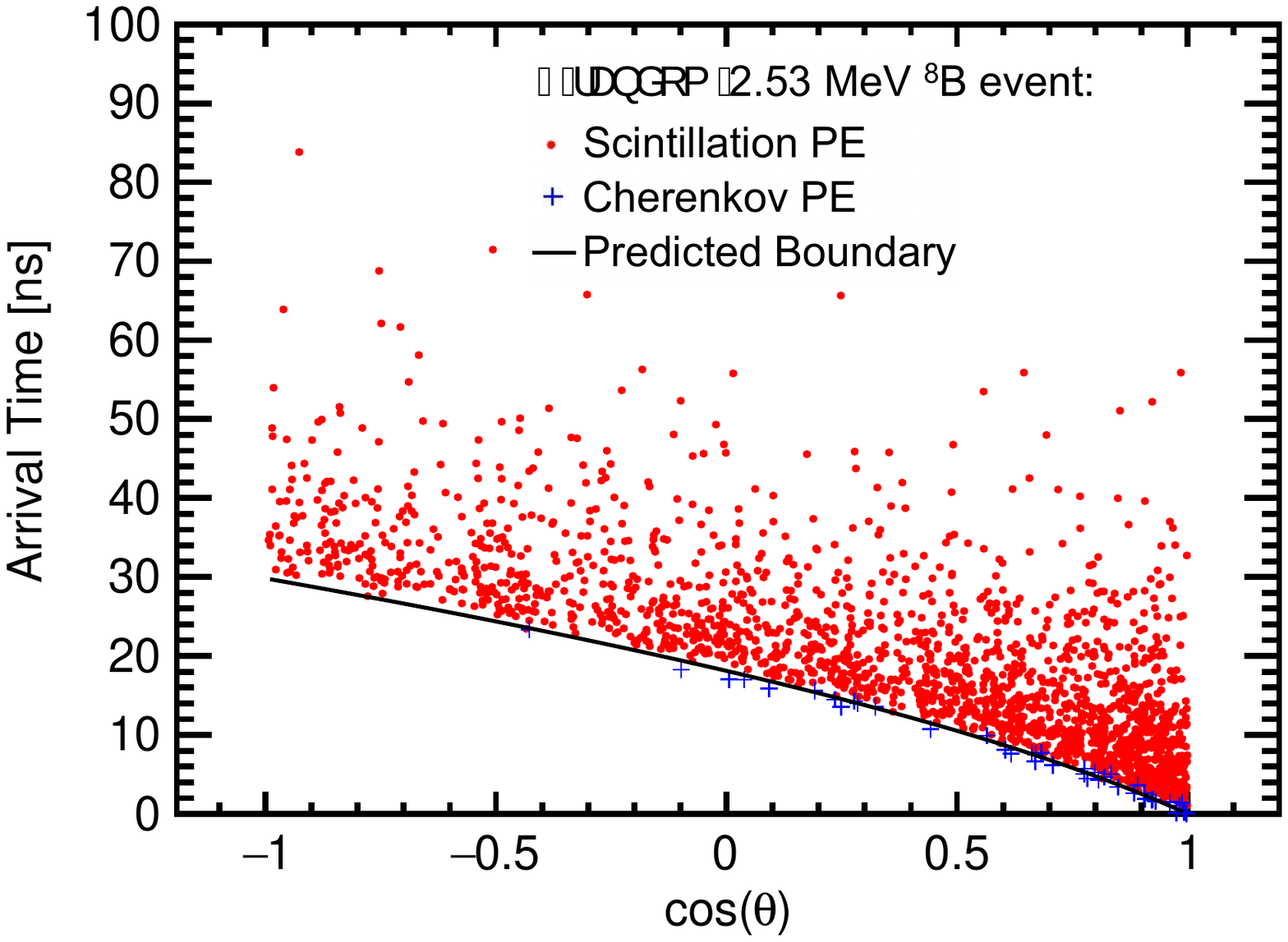}
    \includegraphics[width=0.49\textwidth]{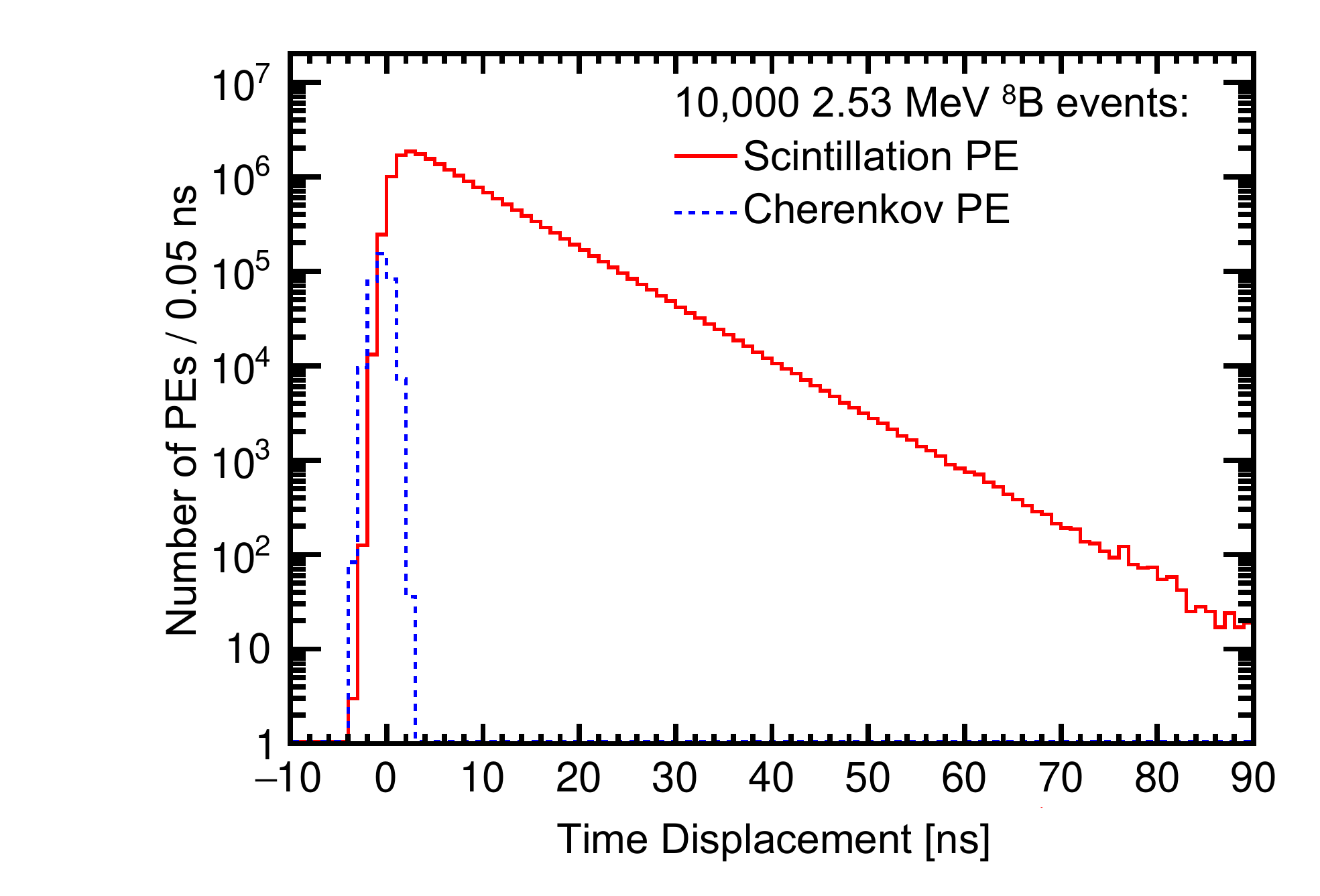}
    \caption{
    [\emph{Left}]: The space-time boundaries $t_c(\theta)$ (solid black line) between Cherenkov PEs (blue crosses) and scintillation PEs (red dots) for one \vbb-decay signal event (top left) and one \B background event (bottom left). The z-axis is chosen to be aligned with the vertex displacement vector $r_v$ from the center of the detector sphere; $\theta$ is defined to be the angle between the z-axis and the photon hit vector $\vec{r}_i$ from the center of the detector sphere.
    The arrival time $t$ is defined relative to the earliest PE detection.
    [\emph{Right}]: Distributions of the time displacement $t_d$ of Cherenkov PEs (dotted blue line) and scintillation PEs (solid red line) from the space-time boundary $t_c(\theta)$, $t_{d}(t_{PE},\theta)=t_{PE}-t_c(\theta)$, for 10,000 \vbb-decay signal events (top right) and 10,000 \B background (bottom right) events. The majority of the scintillation PEs (solid red line) have a larger time displacement $t_d$. All Cherenkov PEs (dotted blue line) are located near the boundary.
    }
    \label{fig:time_boundary_updated}
\end{figure}

Figure~\ref{fig:time_boundary_updated} shows the arrival time of PEs, $t_{PE}$, relative to the very first\footnote{The very first PE arriving to the detector sphere could be either Cherenkov or scintillation depending on the vertex position and electron(s) direction.} PE as a function of the polar angle in one \vbb-decay event and one \B event. One can see that all the Cherenkov PEs are located near the scintillation PE boundary. In the example \vbb-decay signal event shown in Fig.~\ref{fig:time_boundary_updated}, the vertex displacement from the center is $r_v$ = 229~cm and the vertex displacement in the \B background event is $r_v$ = 292~cm.

We define the PEs time displacement, $t_{d}(t_{PE},\theta)$, from the Cherenkov-scintillation space-time boundary as
\begin{eqnarray}
\label{eq:def_time_displacement}
    t_{d}(t_{PE},\theta)=t_{PE}-t_c(\theta).
\end{eqnarray}

The distributions of the time displacement of all PEs at all polar angles in 10,000 signal and 10,000 background events are also shown in Fig.~\ref{fig:time_boundary_updated}. The PE time displacement can be used to select a sample of PEs with higher fraction of Cherenkov PEs compared to a full sample of all PEs in the event.

The deviations from the prediction of the Cherenkov-scintillation space-time boundary given by Eq.~\ref{eq:def_time_boundary} arise due to the uncertainty on the location of the vertex, not knowing the group velocity of individual photons, and not knowing {\it{a priori}} whether the earliest PE is Cherenkov or scintillation.

In the subsequent analysis, we assign a weight $W(t_{PE},\theta)$ to each individual PE based on the time displacement $t_d(t_{PE},\theta)$ in order to increase the contribution from directional Cherenkov light, and to eliminate the effect of the time translation uncertainty of the space-time boundary due to not knowing the type of the earliest PE. In particular, the weight $W(t,\theta)$ is designed such that the total weight on the Cherenkov PEs relative to the scintillation PEs is independent of the time translation uncertainty. We define the weight as

\begin{eqnarray}
\label{eq:def_weight}
    W(t_{PE},\theta)=\exp\left[-\frac{t_{d}(t_{PE},\theta)}{\tau}\right],
\end{eqnarray}
where $\tau$ is a time constant optimized to be 0.4~ns. The chosen value of $\tau = $ 0.4~ns suppresses the contribution of scintillation PEs while still preserves a sufficient contribution from all Cherenkov PEs.

\section{Event Topology Reconstruction}
\label{sec:topology}
\subsection{Rotationally Invariant S-spectrum}
\label{subsec:S-spectrum}
Using the angular coordinates of PEs we construct the following rotationally invariant series, dubbed the `S-spectrum':

\begin{eqnarray}
\label{eq:def_S_spectrum_weight}
	S_{\ell}= \frac{\sum_{m=-\ell}^{\ell} \abs{\sum_{i=1}^{N_{PE}}W(t_i,\theta_i)Y_{\ell m}(\theta_{i}, \phi_{i})}^2}{\abs{\sum_{i=1}^{N_{PE}}W(t_i,\theta_i)}^{2}},
\end{eqnarray}
where $N_{PE}$ is the total number of PEs, $\theta_i$ and $\phi_i$ are the spherical coordinates of each PE and $Y_{\ell m}(\theta,\phi)$ are the tesseral harmonics\footnote{Also known as real-valued spherical harmonics.
See Equation~\ref{EqB:spherical_harmonics} for the definition of tesseral harmonics.}. See~\ref{app:harmonics} for a detailed discussion on the S-spectrum normalization and rotation invariance.

We note that S-spectrum computation time scales linearly with the number of PE hits, $N_{PE}$. In comparison, the computation time of the fast Fourier transform and spherical Fourier transform algorithms on a grid with $N$ points scale as $N \times Log(N)$~\cite{Cooley-Tukey-1965, Kunis-Potts-2003, Rokhlin-Tygert-2006}.

\begin{figure}[hbt!]
    \centering
    \includegraphics[width=0.49\textwidth]{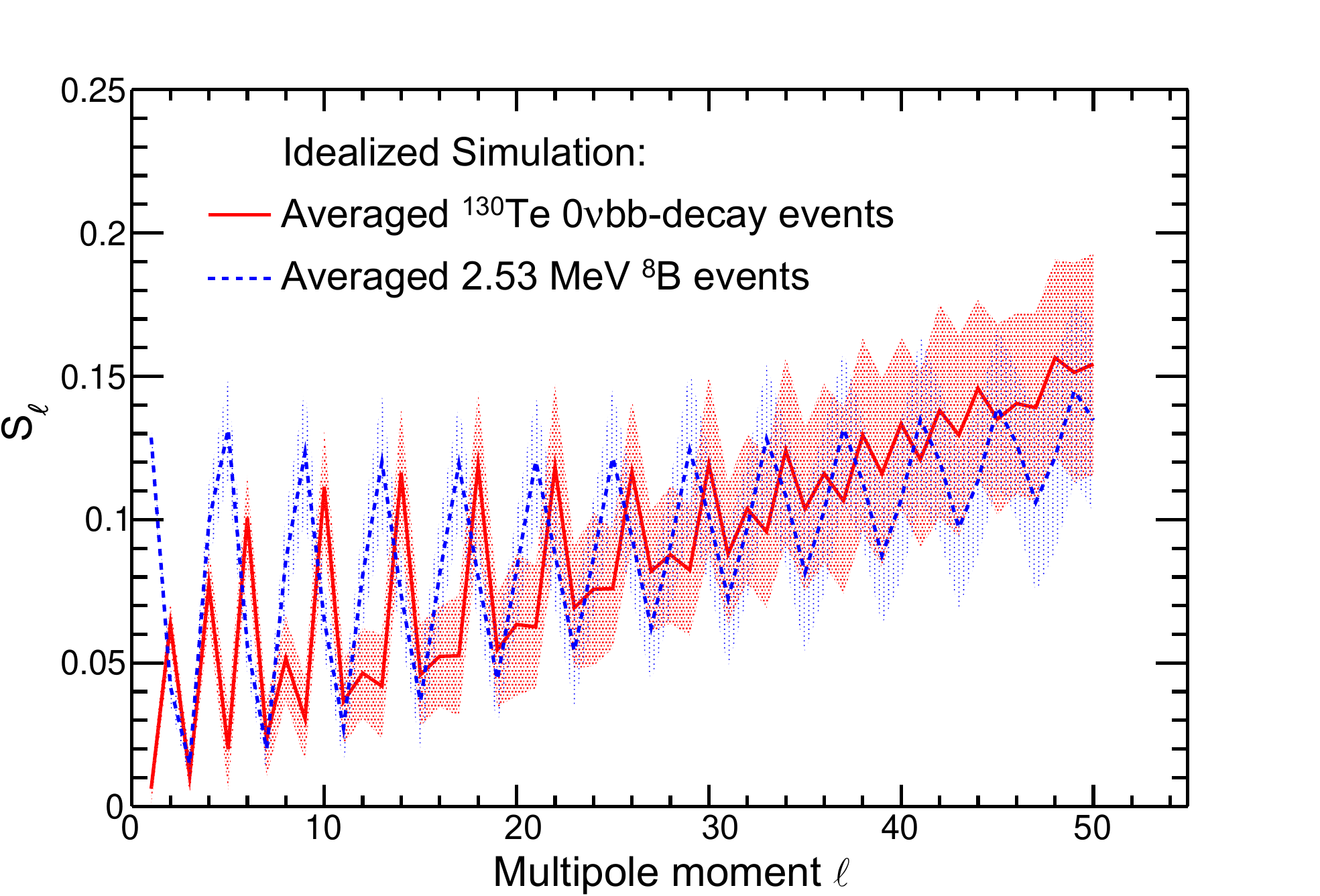}
    \includegraphics[width=0.49\textwidth]{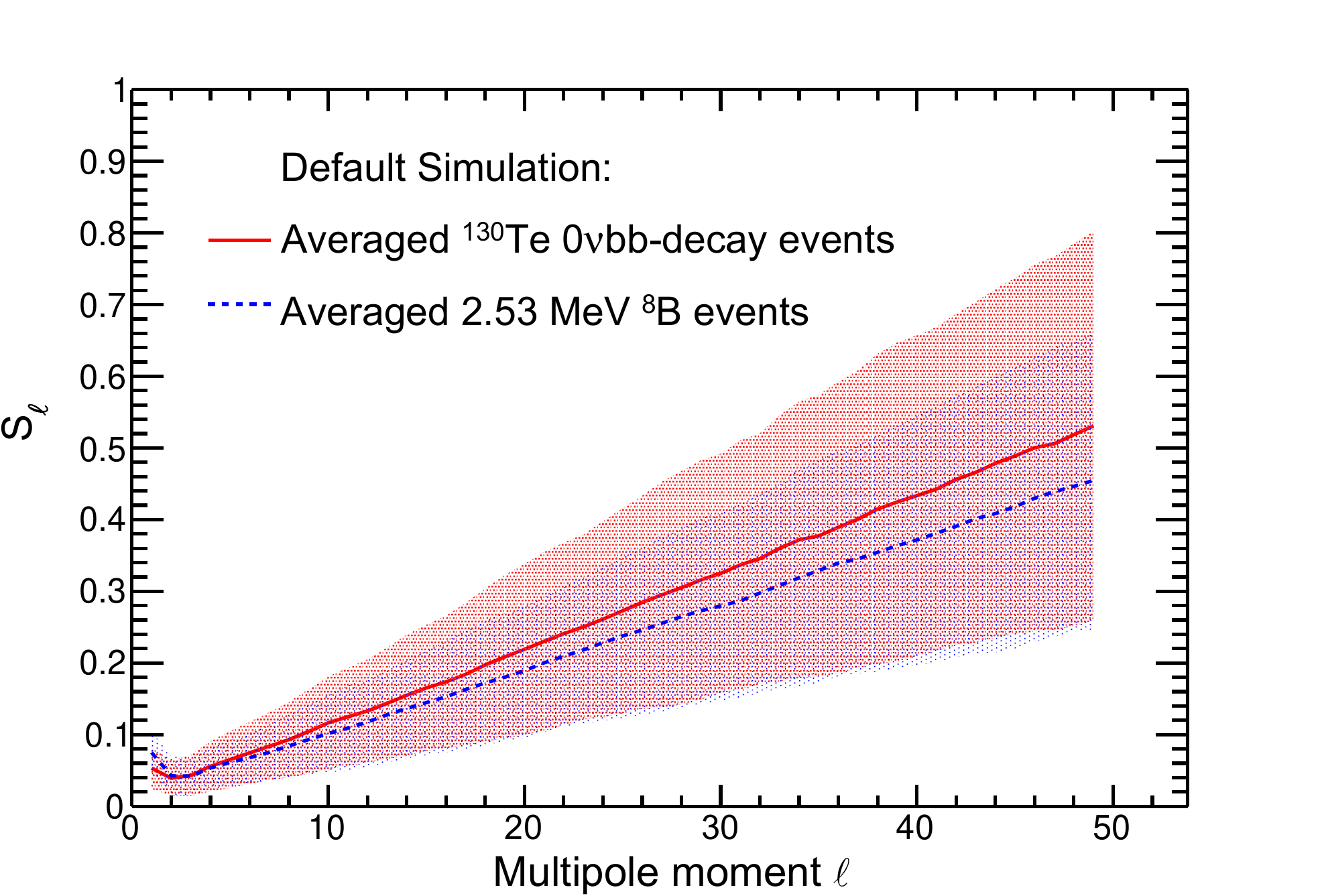}
    \caption{
    The S-spectrum averaged over 100 (left) or 1000 (right) \vbb-decay signal (solid red line) and \B background (dotted blue line) events. Shaded areas show one standard deviation from the mean value indicating a typical event-by-event $S_{\ell}$ spread at each $\ell$.
    [\emph{Left}]: Simulation of 100 events using idealized simulation settings: events originated at the center of the detector sphere, photo-coverage is 100\%, QE is 30\% for both scintillation and Cherenkov light, multiple scattering turned off. To increase the contribution from Cherenkov PEs relative to scintillation PEs, only PEs that arrive within 1.5~ns after the first PE in each event are used for S-spectrum calculations.
    The signal S-spectrum has smaller $S_1$, larger slope $\alpha$, and a more rapid oscillation pattern in the S-spectrum compared to the background S-spectrum.
    [\emph{Right}]: Simulation of 1000 events using default simulation settings: events uniformly distributed throughout the fiducial volume of
    R$<$3~m, photo-coverage is 60\%, QE is 23\% for scintillation light and 12\% for Cherenkov light, multiple scattering is properly included in the simulation. To increase the contribution from Cherenkov PEs relative to scintillation PEs, each PE is assigned weight $W(t_d,\theta)$ according to Eq.~\ref{eq:def_weight}. The signal S-spectrum has smaller $S_1$ and larger slope $\alpha$ compared to the background S-spectrum.
    }
    \label{fig:S_Spectrum_cover_65}
\end{figure}

The S-spectrum series, $S_{\ell}$, is determined by the PE distribution over the sphere and therefore can be used to distinguish different event topologies. Figure~\ref{fig:S_Spectrum_cover_65} shows the S-spectrum averaged over multiple \vbb-decay signal and \B background events for idealized (Fig.~\ref{fig:S_Spectrum_cover_65}, left) and realistic (Fig.~\ref{fig:S_Spectrum_cover_65}, right) simulations. 

In the idealized simulation, events are simulated only at the center of the detector, multiple scattering is turned off, photo-coverage is 100\%, and the QE is 30\% for all photons regardless of the wavelength. Comparing with the background the signal S-spectrum has smaller $S_1$, larger overall slope $\alpha$, and a more rapid oscillation pattern. 

The S-spectrum shape dependence on the photon distribution is discussed in the appendices of Ref.~\cite{Runyu_masters_thesis}. For example, on average, the Cherenkov PE distribution is more symmetric in the case of the two-track 0\nbb ~signal than the one-track \B background, leading to smaller $S_1$ values for the signal.

The idealized simulation is shown here for illustration purposes only. Unless noted otherwise, in this paper we use the default simulation settings as described in Sec.~\ref{sec:detector_simulation}. In the default simulation, multiple scattering is included, photo-coverage is 65\%, QE is of a typical bialkali photo-cathode, and all simulated events are uniformly distributed in the detector fiducial volume of R$<$3~m. 

As shown in Fig.~\ref{fig:S_Spectrum_cover_65} (right) for events simulated with the default simulation settings the main differences between signal and background S-spectra are in the values of $S_1$ and the slope $\alpha$. Therefore we use the $S_1$ and $\alpha$ parameters to separate \vbb-decay signal from \B background events.

\begin{figure}[hbt!]
    \includegraphics[width=0.49\textwidth]{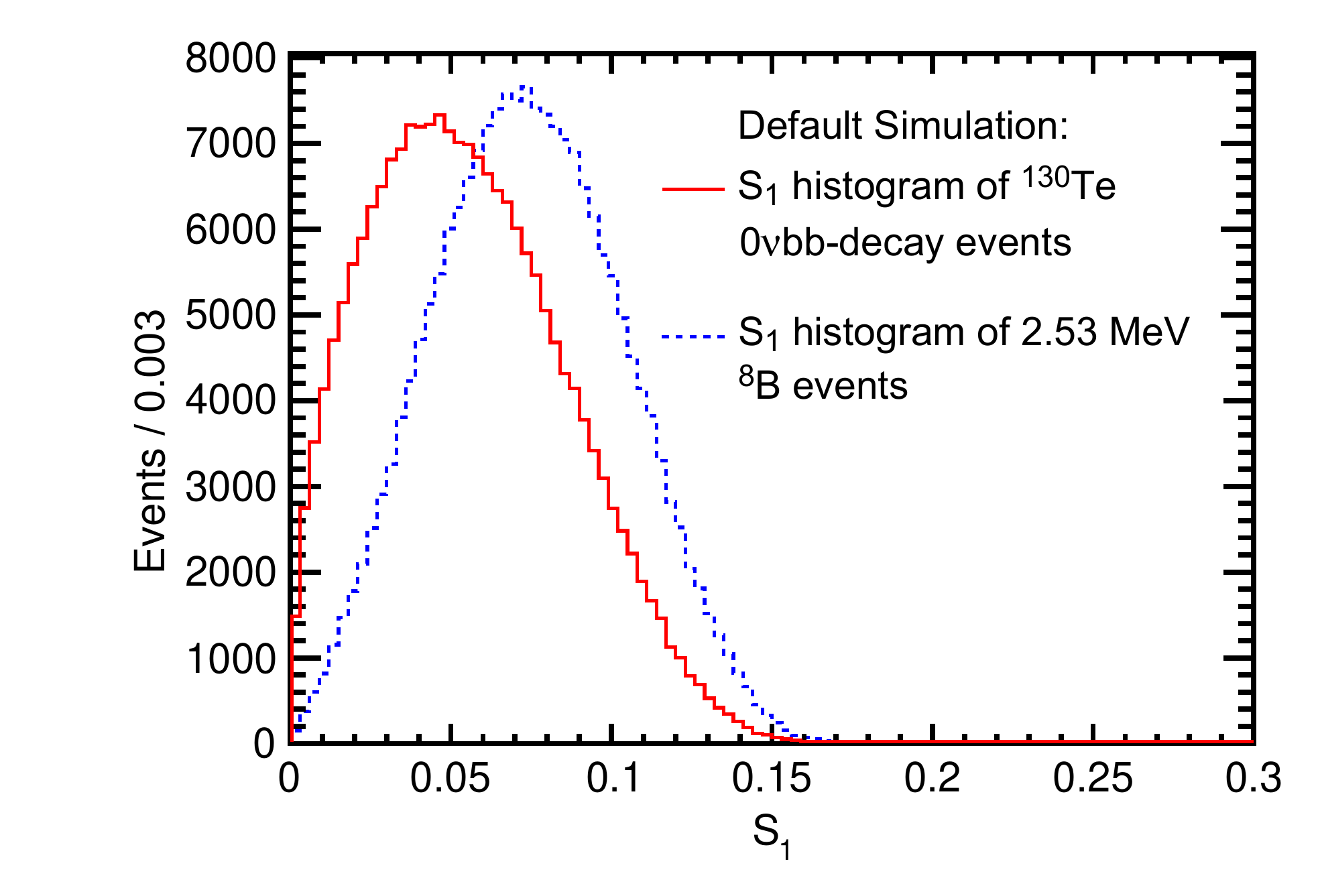}
    \includegraphics[width=0.49\textwidth]{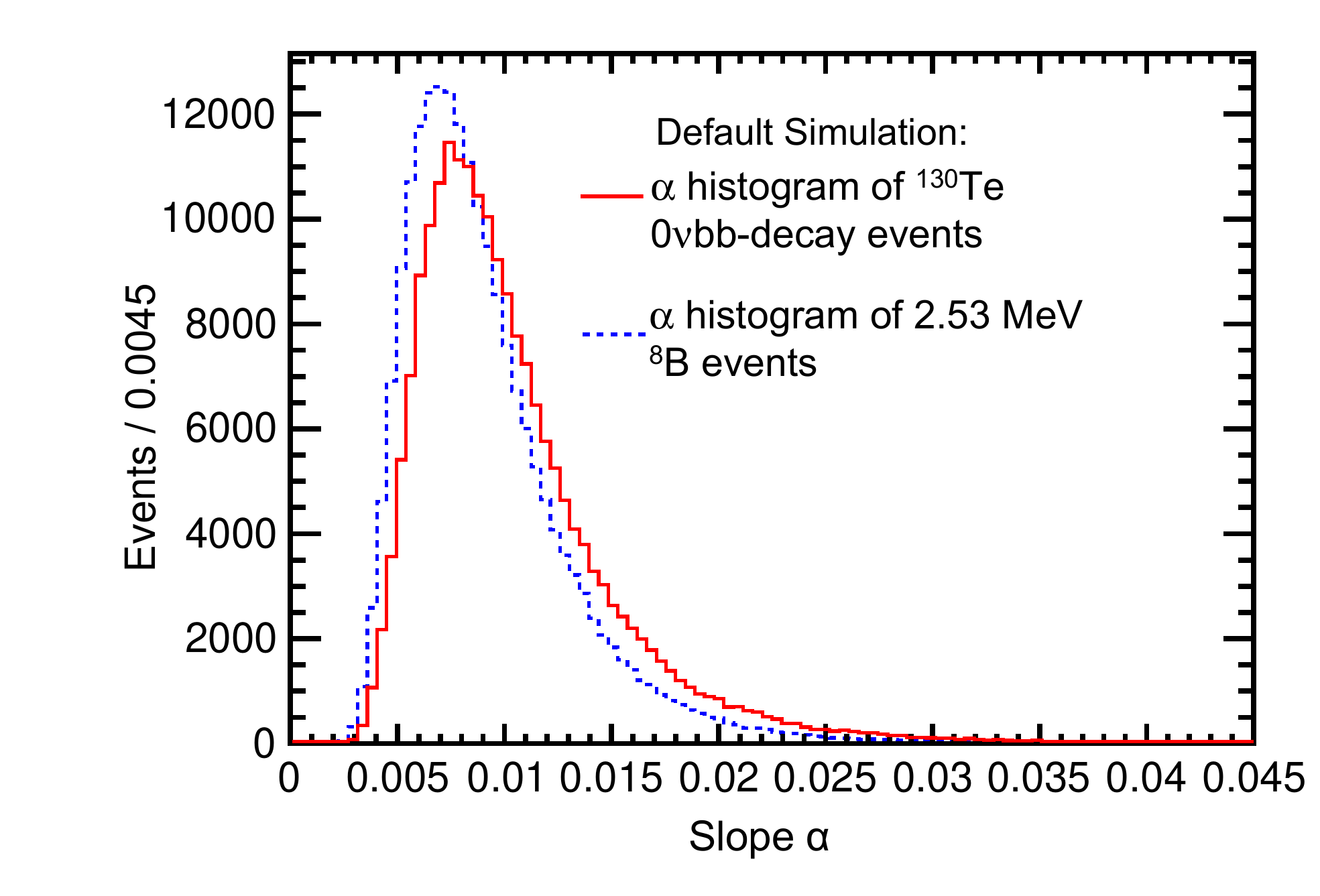}
    \caption{
    [\emph{Left}]: Comparison of the $S_1$ parameter between 200,000 simulated 0\nbb ~signal (solid red line) and \B background (dotted blue line) events. The default simulation settings are used.
    [\emph{Right}]: Comparison of the $\alpha$ parameter between 200,000 0\nbb ~signal (solid red line) and \B background (dotted blue line) events. 
    }
    \label{fig:Parameter_Space}
\end{figure}

The distribution of $S_1$ and $\alpha$ parameters for \vbb-decay signal and \B background events are shown in Fig.~\ref{fig:Parameter_Space}. Using these two distributions we construct a likelihood function to separate signal and background events.

\subsection{Maximum Likelihood Estimation}

For a candidate event in the fiducial volume, the values of parameters $S_1$ and $\alpha$ constitute a measurement, $\vec{S}_{meas}=(S_1,\alpha)$. The likelihood, $p(2e|\vec{S}_{meas})$, for such an event to be a 0\nbb-decay signal is given by:

\begin{eqnarray}
\label{eq:def_likelihood}
    p(2e|\vec{S}_{meas})=\frac{p(\vec{S}_{meas}|2e)}{p(\vec{S}_{meas}|1e)+p(\vec{S}_{meas}|2e)}=\frac{p(S_1|2e)p(\alpha|2e)}{\sum_{j=1}^{2}p(S_1|je)p(\alpha|je)},
\end{eqnarray}
where $2e$ is a label for the two-track signal events, $1e$ is the label for the one-track background events,
$p(S_1|2e)$ and $p(\alpha|2e)$ are the probability distributions of $S_1$ and $\alpha$ in 0\nbb~signal events respectively; similarly, $p(S_1|1e)$ and $p(\alpha|1e)$ are the probability distribution of $S_1$ and $\alpha$ in \B background events respectively. Note that we make an approximation by replacing the two-dimensional probability distribution by the product of two one-dimensional probability distributions (see~\ref{app:likelihood} for details).

\begin{figure}[hbt!]
    \includegraphics[width=0.49\textwidth]{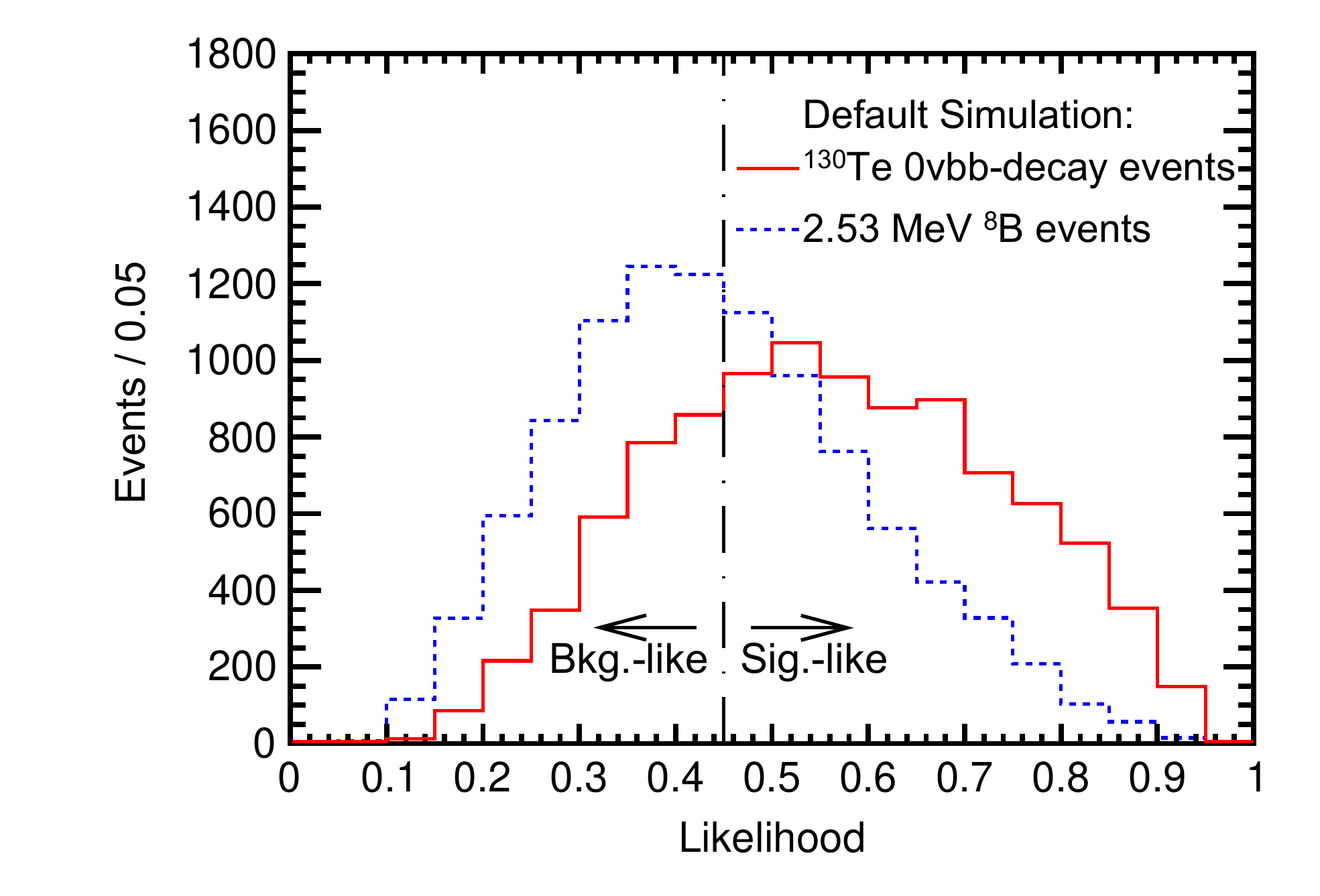}
    \includegraphics[width=0.49\textwidth]{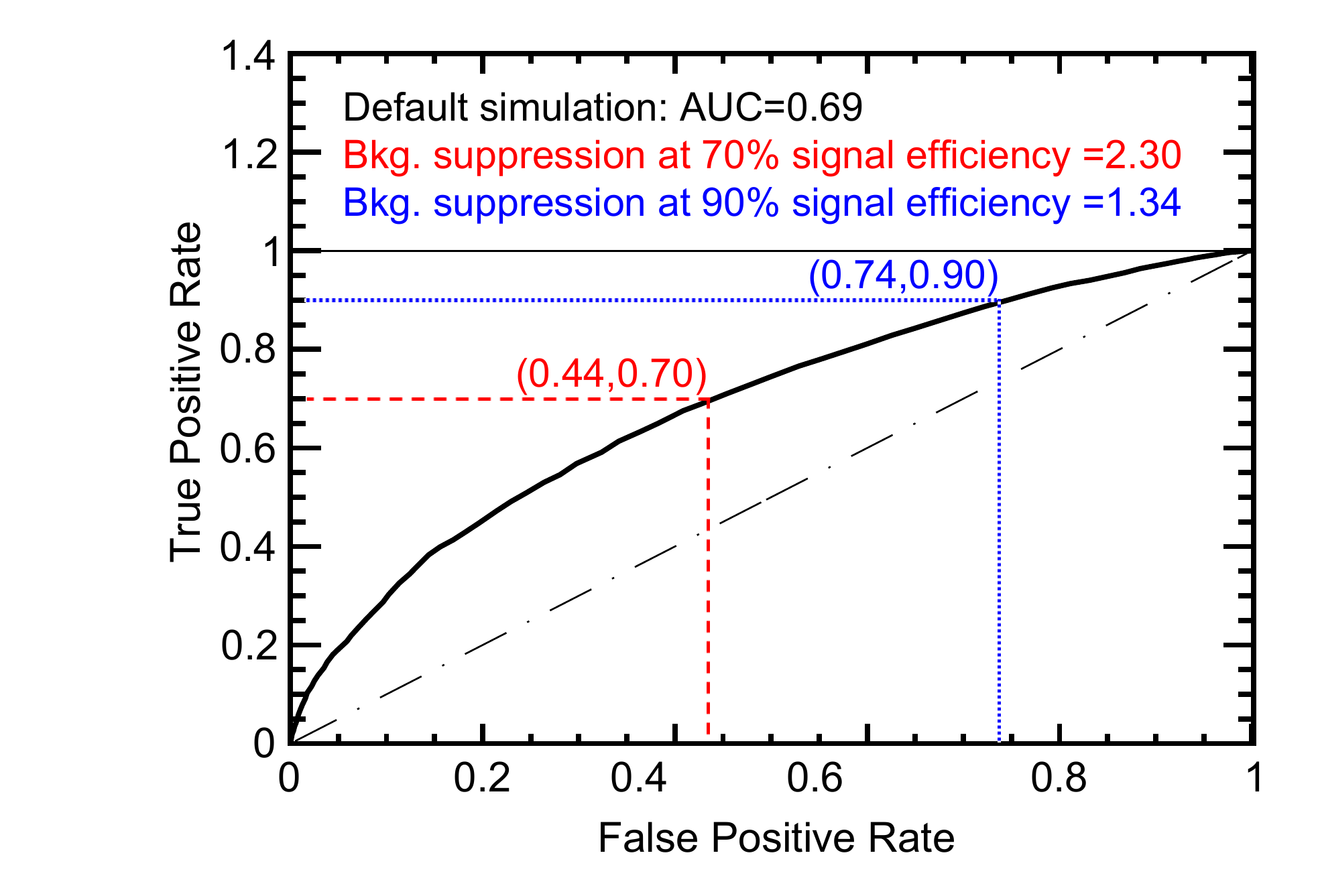}
    \caption{
    [\emph{Left}]: The distribution of the likelihood values, $L$, calculated using Eq.~\ref{eq:def_likelihood} for 10,000 simulated 0\nbb ~signal (solid red line) and \B background (dotted blue line) events. The vertical black dash-dotted line at 0.45 corresponds to a cut on the likelihood value that results in 70\% signal efficiency. The default simulation settings are used. 
    [\emph{Right}]: The true positive rate $TPR(L_{cut})$ as a function of the false positive rate $FPR(L_{cut})$, or the ROC-curve. Using the ROC-curve we determine that the background suppression factor is 2.30 at 70\% signal efficiency (dashed red lines) and 1.34 at 90\% signal efficiency (dotted blue lines). The AUC for this ROC-curve is 0.69. The horizontal black thin solid line at $TPR = $1 corresponds to an ideal event classification algorithm with 100\% separation between signal and background (AUC=1); the black dash-dotted diagonal line corresponds to a classification algorithm based on a random guess with no separation between signal and background (AUC=0.5).
    }
    \label{fig:likelihood_and_ROC}
\end{figure}

Figure~\ref{fig:likelihood_and_ROC} (left) shows the likelihood distribution $p(2e|\vec{S}_{meas})$ for 10,000 signal and background events. The likelihood distribution for the signal events is shifted to the right with respect to the likelihood distribution for background events. We use this difference in the subsequent analysis to separate signal and background events.

\subsection{Event Classification Results}

Introducing a likelihood threshold, $L_{cut}$, in the likelihood distribution $p(2e|\vec{S}_{meas})$, we can classify an event with a likelihood $p(2e|\vec{S}_{meas}) > L_{cut}$ to be a signal event.
We scan over $L_{cut}$ values to determine the background suppression factors at a signal efficiencies of 70\% and 90\%.
For example, for the likelihood distribution shown on the left-hand panel of Fig.~\ref{fig:likelihood_and_ROC}, 70\% signal
efficiency corresponds to the $L_{cut}$ value of 0.45. Rejecting events with $p(2e|\vec{S}_{meas}) < 0.45$ results in rejecting 
56.5\% of the simulated \B background events, {\it i.e.} a background suppression factor of 2.3.

For any given value of the likelihood threshold $L_{cut}$, we define the true positive rate (TPR) as a fraction of signal events that satisfy $p(2e|\vec{S}_{meas}) > L_{cut}$. In other words, the TPR is a fraction of signal events that are correctly classified as signal. The TPR is the signal efficiency.

Similarly, we define the false positive rate (FPR) as a fraction of background events that satisfy $p(2e|\vec{S}_{meas}) > L_{cut}$. In other words, the FPR is a fraction of background events that are falsely classified as signal. The inverse value of the FPR, $\frac{1}{FPR}$, is the background suppression factor.

The true positive rate (TPR) as a function of the false positive rate (FPR) is known as the receiver operating characteristic curve (ROC-curve). 
The ROC-curve is commonly used in computer science to characterize the performance of a classification algorithm.

The right-hand panel on Fig.~\ref{fig:likelihood_and_ROC} shows the ROC-curve, $TPR = f(FPR)$, corresponding to the topological 
reconstruction for the default simulation. For comparison, the ROC-curve of a perfect reconstruction is shown as a 
horizontal line at $TPR = 1$. The ROC-curve of a classification algorithm based on a random guess is shown as a diagonal line, $TPR = FPR$.

We use the area under the ROC-curve (AUC) as a figure of merit to numerically characterize the performance of the topological reconstruction.
The AUC of 0.5 corresponds to no separation between signal and background. The AUC of 1 corresponds to a perfect separation between signal and background. For the topological reconstruction shown in Fig.~\ref{fig:likelihood_and_ROC} (right) the AUC is 0.69. The background suppression factor is 2.30 and 1.34 at 70\% and 90\% signal efficiency respectively.

\begin{figure}[hbt!]
    \centering
    \includegraphics[width=0.49\textwidth]{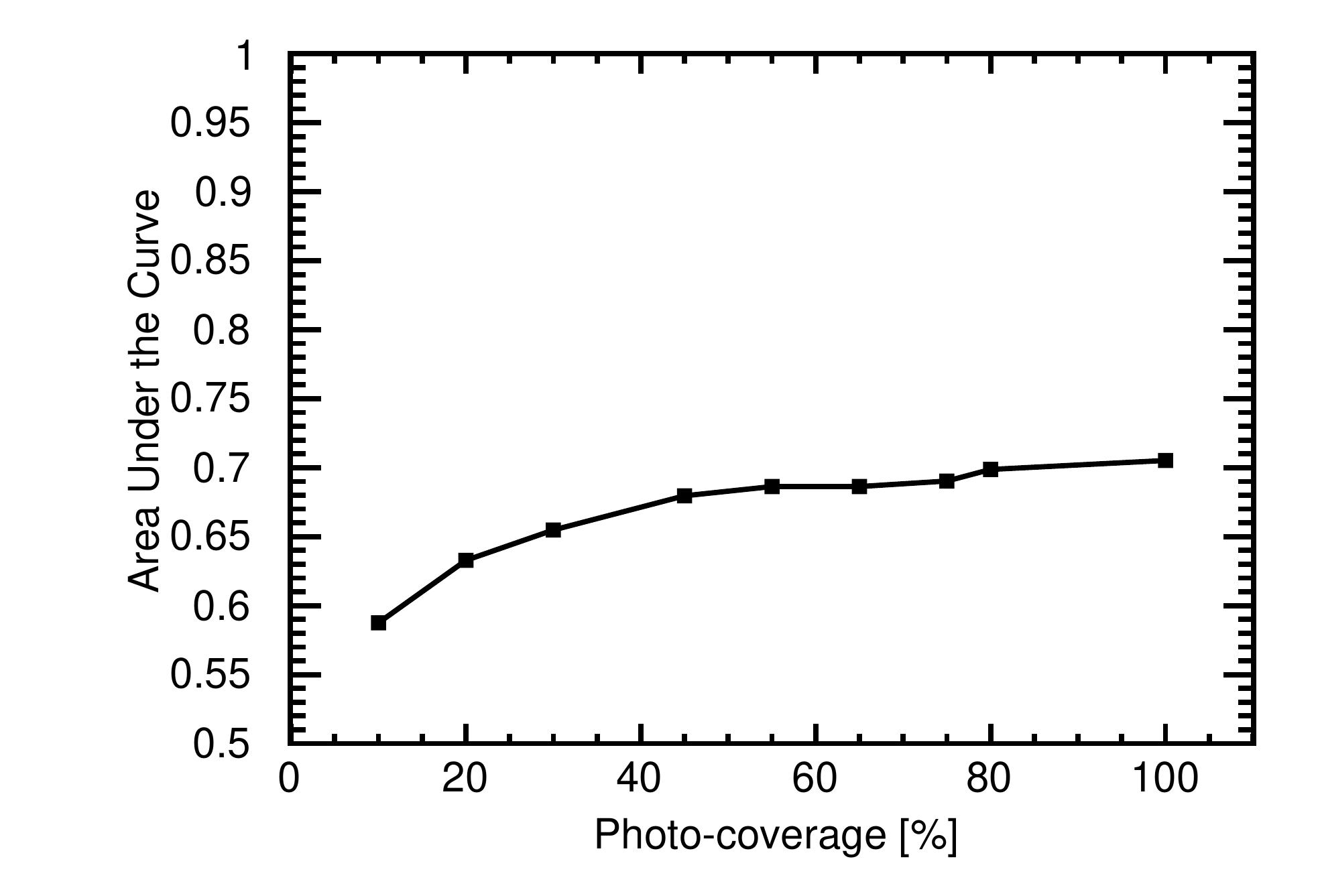}
    \includegraphics[width=0.49\textwidth]{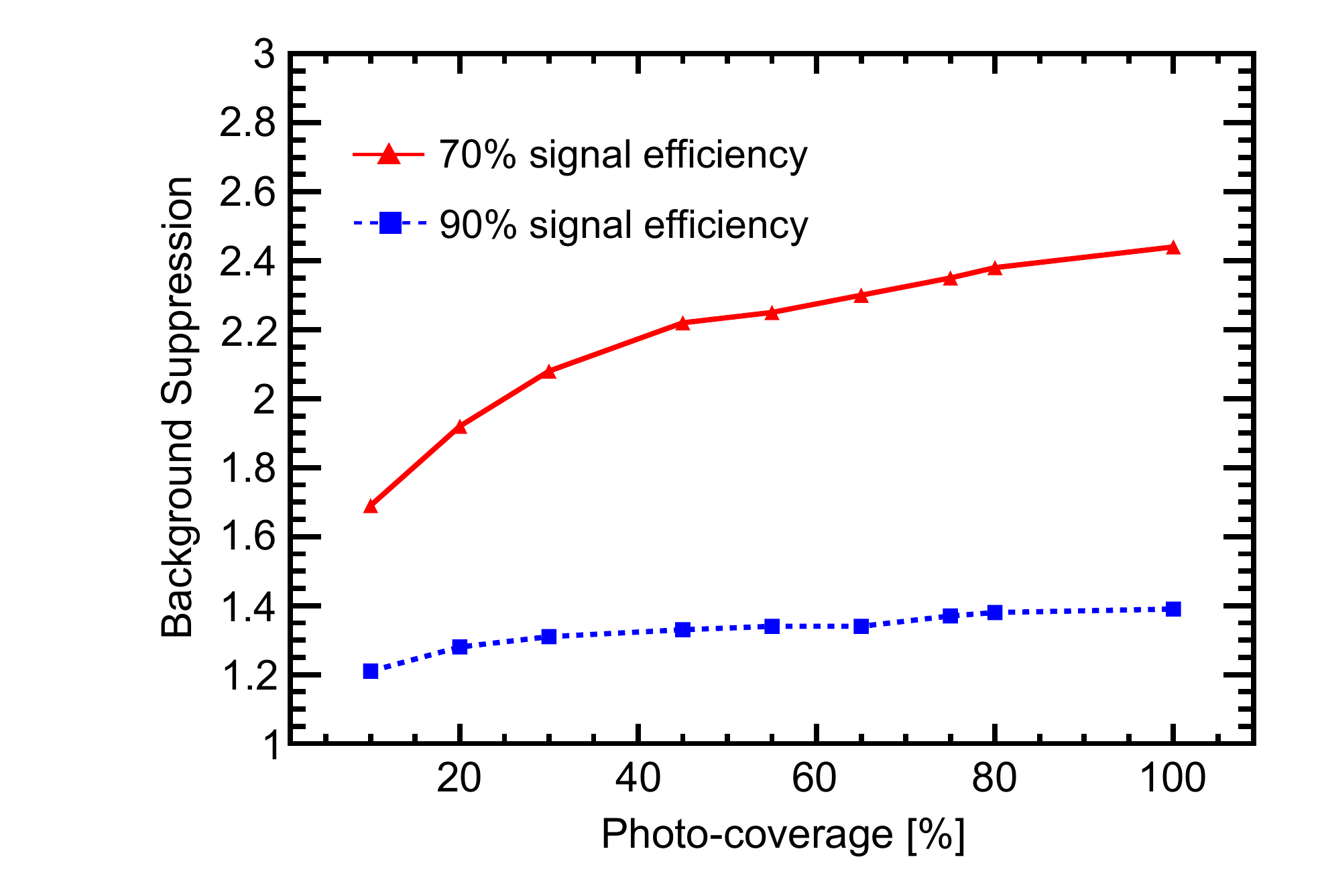}
    \caption{[\emph{Left}]: The area under the curve (AUC) as a function of the photo-coverage. Each point is calculated based on the likelihood distributions of 10,000 0\nbb ~signal and \B background events simulated with all the  default simulation settings except for the photo-coverage, which is varied from 10\% to 100\%.  
    [\emph{Right}]: The background suppression factors at 70\% signal efficiency (solid red line) and 90\% signal efficiency (dotted blue line) as a function of the vertex smearing. Each point is calculated based on the ROC curve of 10,000 0\nbb ~signal and \B background events simulated with all the  default simulation settings except for the vertex smearing, which is varied from 10\% to 100\%.
    }
    \label{fig:ROC_vs_coverage}
\end{figure}

Figure~\ref{fig:ROC_vs_coverage} shows the dependence of the topological reconstruction on the photo-coverage. We vary the photo-coverage
in the default simulation from 10\% to 100\%. For 100\% coverage the AUC is 0.71 and the background suppression factors are 1.38 and 2.44 at 
90\% and 70\% signal efficiency respectively. At 10\% coverage the AUC and the background suppression factors drop to 0.59, 1.21, and 1.69, respectively.

We note that the default simulation with 100\% photo-coverage matches the simulation settings used in the previously published 
spherical harmonics analysis~\cite{harmonics2017}. Therefore, the background suppression factor of 2.44 can be directly compared 
with the factor of about 2.0 for the technique described in Ref.~\cite{harmonics2017}~\footnote{For a similar comparison, 
in Ref.~\cite{Runyu_masters_thesis} the corresponding background suppression factor is quoted to be 2.6, which is a rounded off 2.56. The
5\% difference between 2.56 in Ref.~\cite{Runyu_masters_thesis} and 2.44 quoted in this paper is due to a statistical fluctuation. 
The data set that we use here for the ROC-curve calculations is 10 times larger than the the data set in Ref.~\cite{Runyu_masters_thesis}.}. 
The topological reconstruction presented here leads to a more than 20\% improvement in background suppression at 70\% signal efficiency.

\begin{figure}[hbt!]
    \centering
    \includegraphics[width=0.49\textwidth]{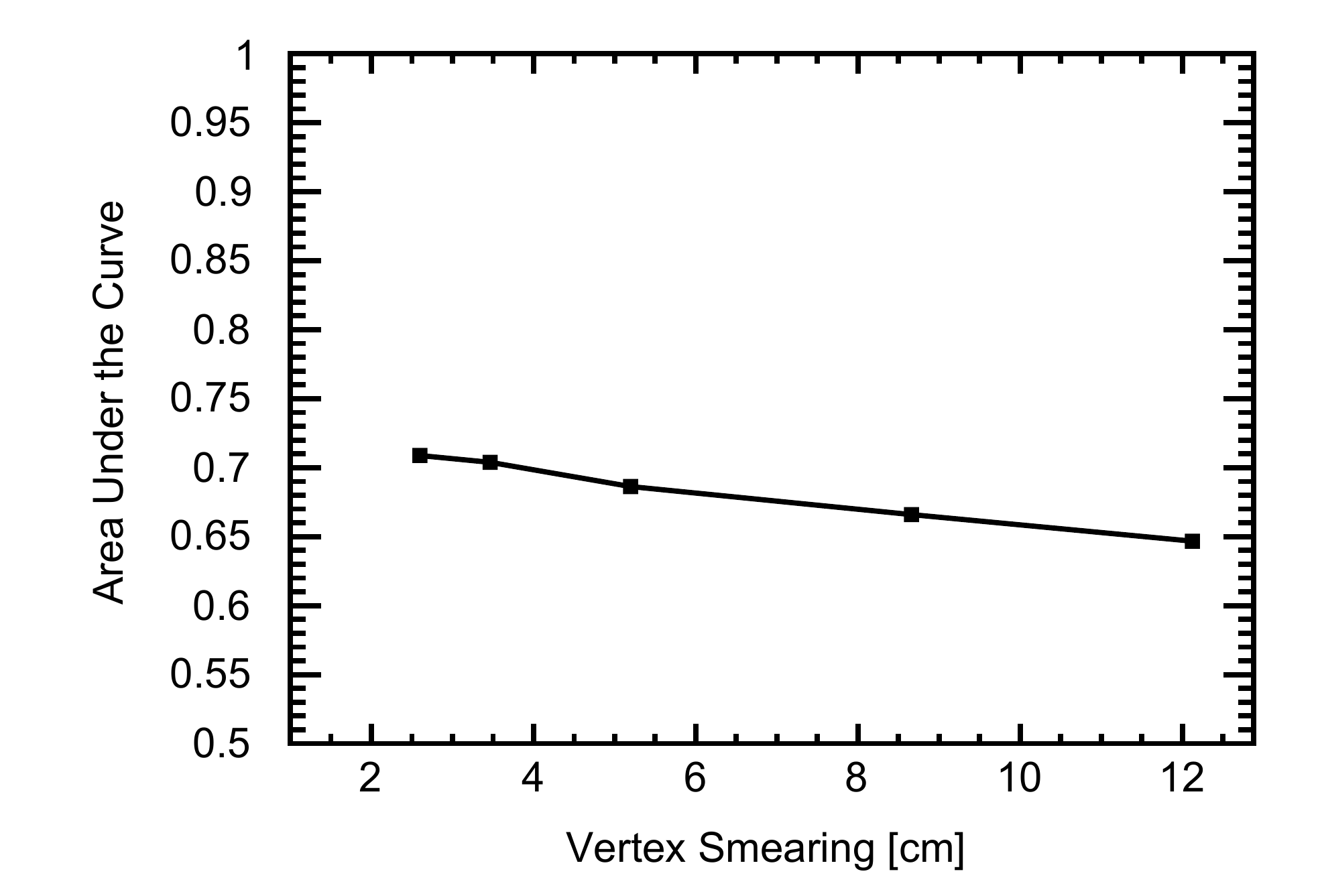}
    \includegraphics[width=0.49\textwidth]{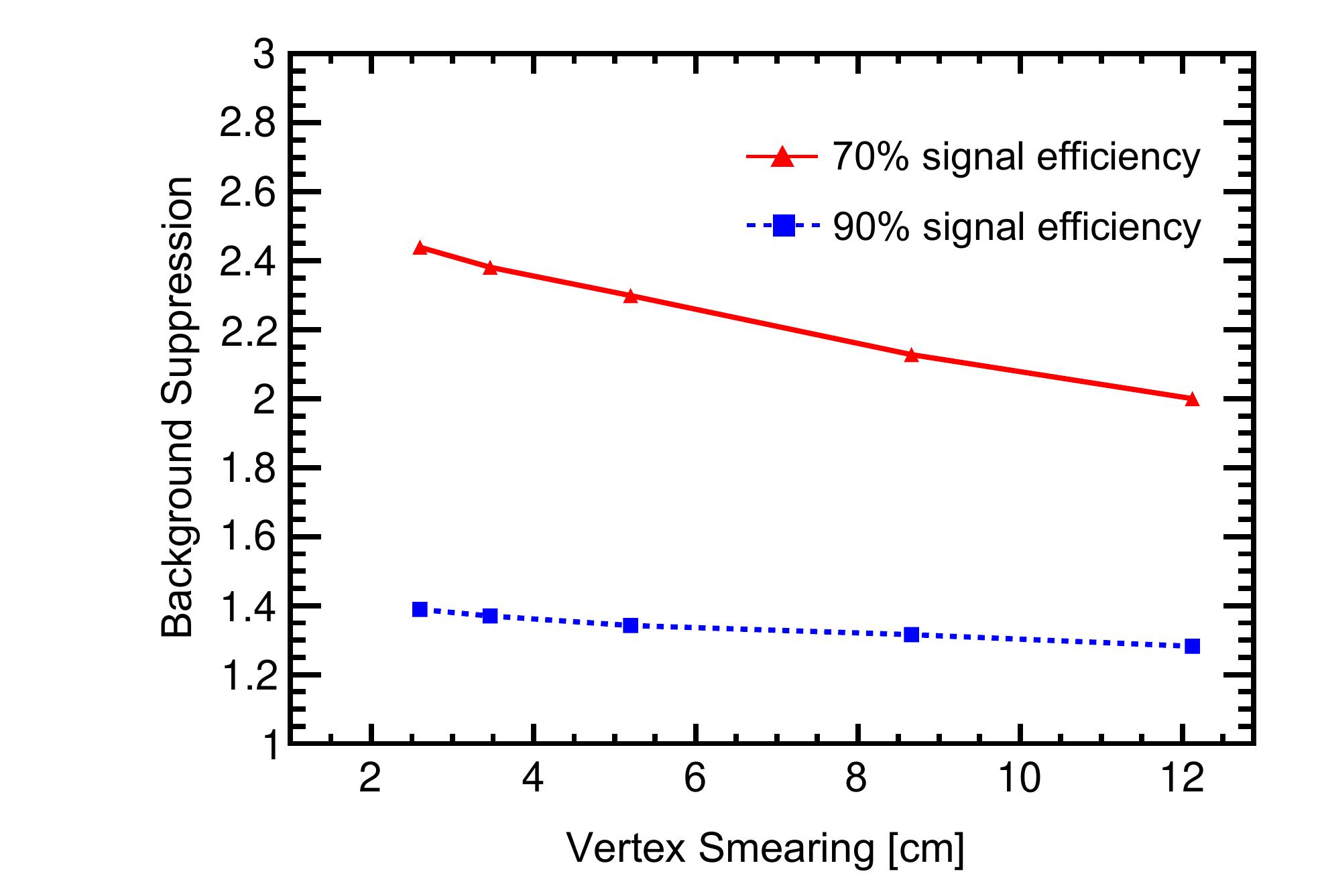}
    \caption{[\emph{Left}]: The area under the curve (AUC) as a function of the vertex smearing, $\sigma_{r_v}$. Each point is calculated based on the likelihood distributions of 10,000 0\nbb ~signal and \B background events simulated with all the  default simulation settings except for the vertex smearing, which is varied from 2.6 cm to 12.1 cm (this corresponds to 1.5-7 cm range in $\sigma_{x,y,z}$).
    [\emph{Right}]: The background suppression with 70\% signal efficiency (solid red line) and 90\% signal efficiency (dotted blue line) as a function of the vertex smearing, $\sigma_{r_v}$. Each point is calculated based on the ROC curve of 10,000 0\nbb ~signal and \B background events simulated with all the  default simulation settings except for the vertex smearing, which is varied from 2.6 cm to 12.1 cm (this corresponds to 1.5-7 cm range in $\sigma_{x,y,z}$).
    }
    \label{fig:ROC_smear}
\end{figure}

We also considered a detector simulation with an ideal light collection: both photo-coverage and QE are 100\%, while all other parameters are the same as in the default simulation. For this scenario, we find the background suppression factors of 1.56 and 3.57 at 90\% and 70\% signal efficiency respectively.

We find a similar performance of the topological reconstruction for a scenario where we change the photo-coverage back to the default 65\%, remove all scintillation light, and filter Cherenkov PEs through a uniform QE of 30\%. For this scenario we find the background suppression factors of 1.59 and 3.39 at 90\% and 70\% signal efficiency respectively. This emphasizes the importance of further developments in techniques to separate Cherenkov PEs from scintillation PEs.

We also study the dependence of the topological reconstruction on the vertex resolution. Using the default simulation with 65\% photo-coverage
we vary the vertex smearing from $\sigma_{r_v}$ = 2.6~cm ($\sigma_{x,y,z}$ = 1.5~cm) to $\sigma_{r_v}$ = 12.1~cm ($\sigma_{x,y,z}$ = 7~cm).
Figure~\ref{fig:ROC_smear} shows the AUC and the background suppression factors as a function of vertex smearing. Improvements in the
vertex resolution lead to a better performance of the topological reconstruction.

\section{Directionality Reconstruction}
\label{sec:directionality}
The topological reconstruction discussed in Sec.~\ref{sec:topology} cannot distinguish signal and background events if in signal events the two electrons are emitted at a small angle or if one electron is emitted with kinetic energy below Cherenkov threshold. However, electron directionality provides an extra handle to separate \B background from \vbb-decay signal events that are misidentified as one-track events. The direction of the electron in a \B background event is correlated with the direction to the Sun, while the directions of the electrons in a 0\nbb ~signal event do not have such correlation. Therefore, \B background can be further suppressed by reconstructing the directionality of one-track candidate events.

We reconstruct the direction of the electron $\hat{r}_e$ using the average of the weighted unit hit vector $\frac{\vec{r}_i-\vec{r}_v}{\abs{\vec{r}_i-\vec{r}_v}}$ of the PEs from the vertex $\vec{r}_v$:
\begin{eqnarray}
\label{eq:electron_direction}
\hat{r}_e=\frac{\sum_{i=1}^{N_{PE}}W(t_i,\theta_i)\frac{\vec{r}_i-\vec{r}_v}{\abs{\vec{r}_i-\vec{r}_v}}}{\abs{\sum_{i=1}^{N_{PE}}W(t_i,\theta_i)\frac{\vec{r}_i-\vec{r}_v}{\abs{\vec{r}_i-\vec{r}_v}}}}
\end{eqnarray}

\begin{figure}[hbt!]
    \centering
    \includegraphics[width=0.49\textwidth]{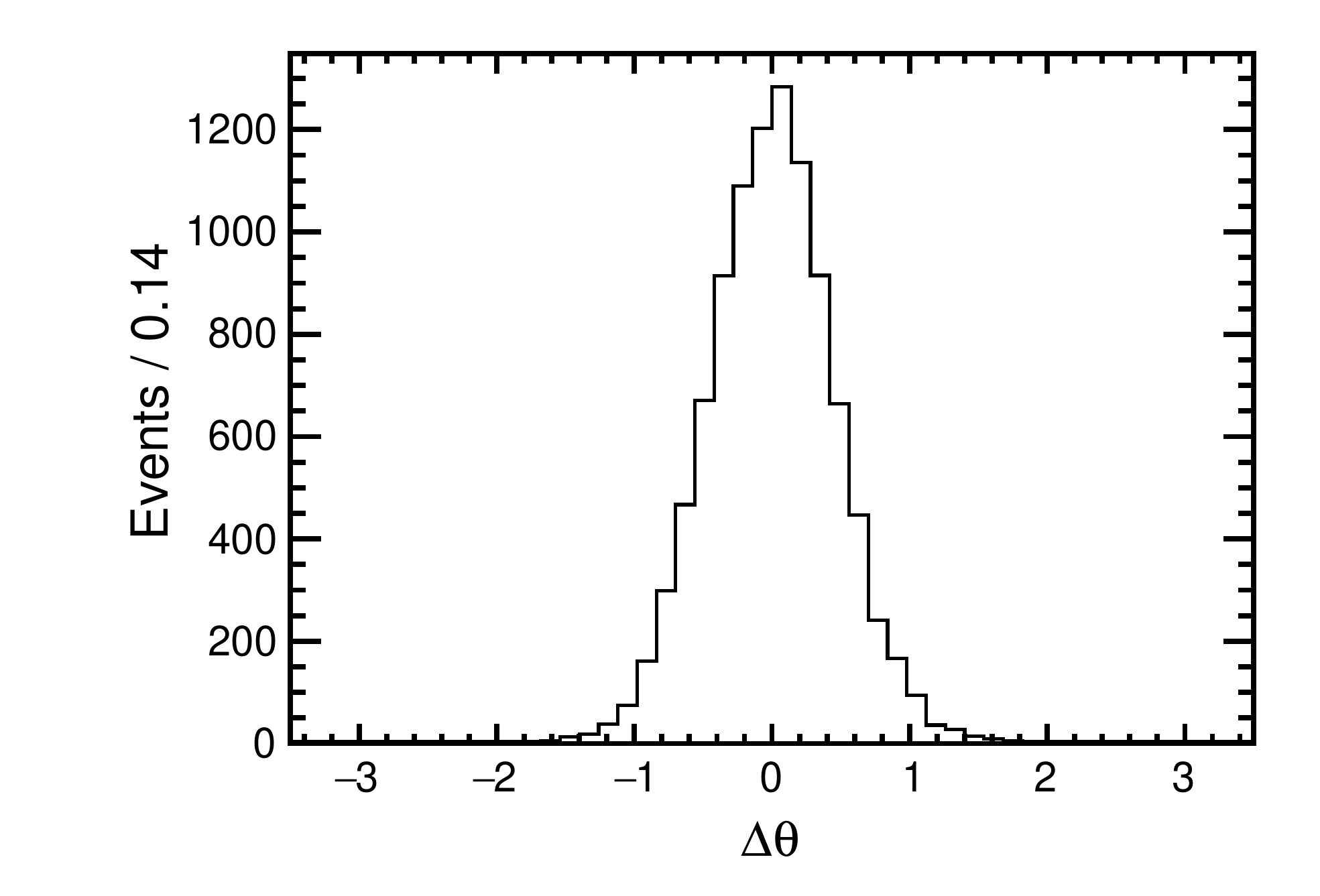}
    \includegraphics[width=0.49\textwidth]{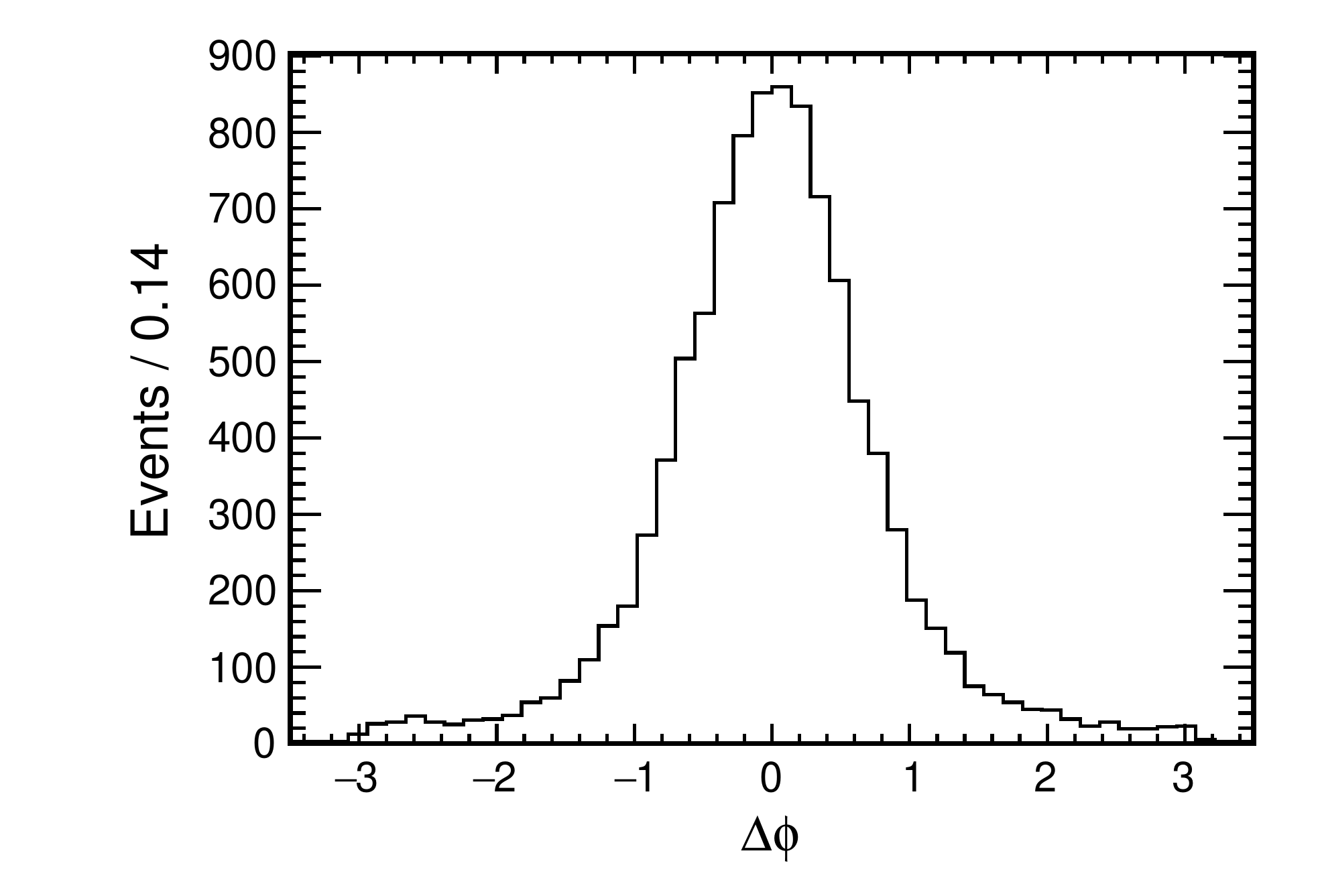}
    \caption{
    Comparison between reconstructed and true direction of the electron $\hat{r}_e$. Simulation of 10,000 one-track \B background events with the default simulation settings.
    [\emph{Left}]: Difference between reconstructed and true polar angle $\Delta \theta = \theta_{reco} - \theta_{true}$. The RMS value of the $\Delta \theta$ distribution is 0.46. 
    [\emph{Right}]: Difference between reconstructed and true azimuthal angle $\Delta \phi = \phi_{reco} - \phi_{true}$. The RMS value of the $\Delta \phi$ distribution is 0.84. 
    }
    \label{fig:theta_and_phi_deviation_vs_coverage}
\end{figure}

Figure~\ref{fig:theta_and_phi_deviation_vs_coverage} shows the comparison between the reconstructed and the true direction of the electron. The distribution of the difference between the reconstructed and the true polar angle, $\Delta \theta = \theta_{reco} - \theta_{true}$, has an RMS value of 0.46. The distribution of the difference between the reconstructed and the true azimuthal angle, $\Delta \phi = \phi_{reco} - \phi_{true}$, has an RMS value of 0.84.

\begin{figure}[h!]
    \centering
    \includegraphics[width=0.49\textwidth]{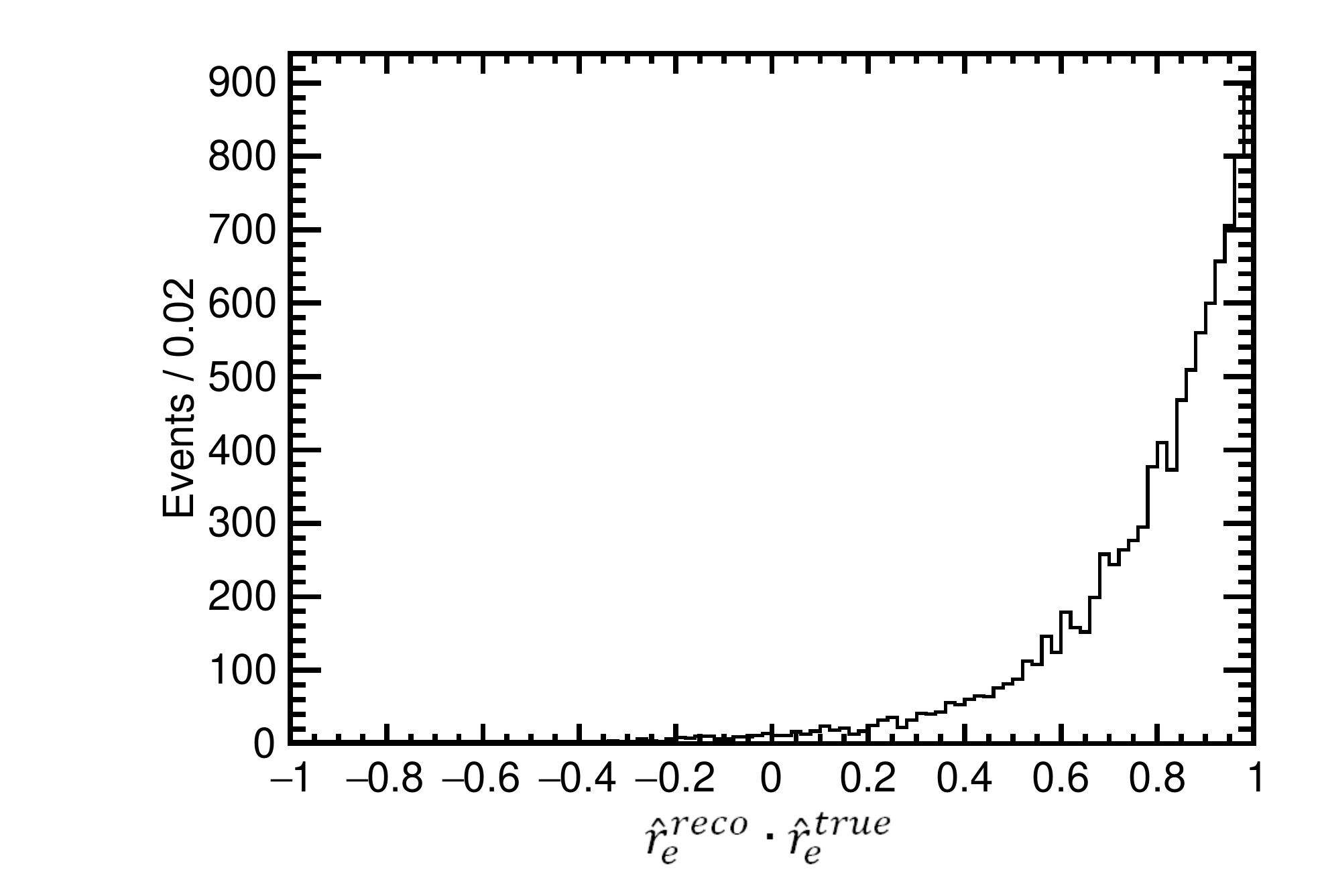}
    \includegraphics[width=0.49\textwidth]{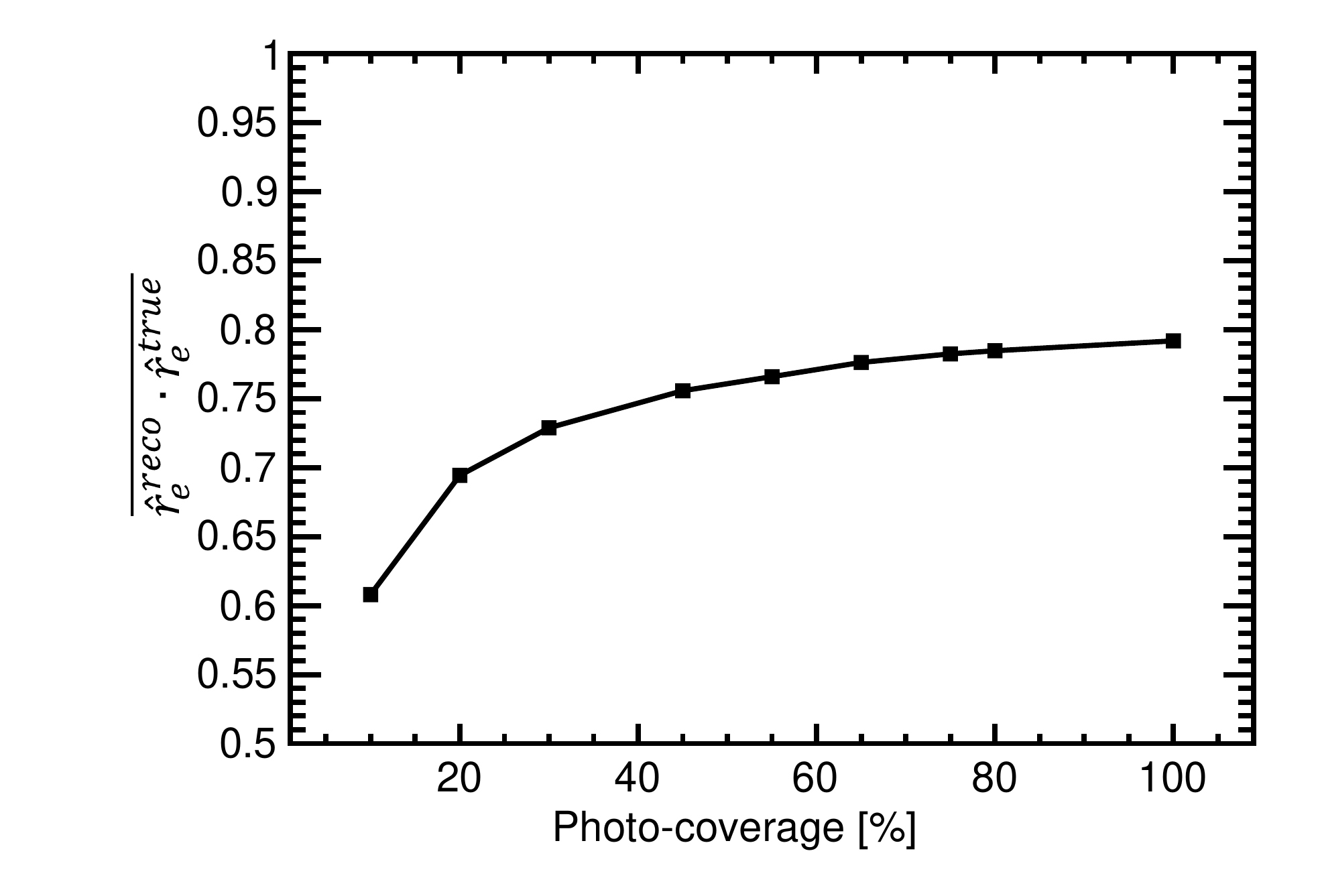}
    \caption{
    [\emph{Left}]: The inner product between the reconstructed direction and the true electron direction for simulation of 10,000 one-track \B background events with the default simulation settings.
    [\emph{Right}]: The mean value of the inner product distribution as a function of photo-coverage.
    The default simulation settings are used except for the photo-coverage, which is varied from 10\% to 100\%.
    }
    \label{fig:inner_product_coverage}
\end{figure}

As in Ref.~\cite{Aberle2014} we also use the inner product between the reconstructed electron direction and the true electron direction as a figure of merit for directionality reconstruction. The left-hand panel in Fig.~\ref{fig:inner_product_coverage} shows the inner product between the reconstructed electron direction and the true electron direction. The mean value of the distribution shown in Fig.~\ref{fig:inner_product_coverage} (left) is 0.78, where the mean inner product of 1 would correspond to perfect directionality reconstruction.

The right-hand panel in Fig.~\ref{fig:inner_product_coverage} shows the mean value of the inner product distribution as a function of photo-coverage. As expected, the inner product increases with larger photo-coverage.

To estimate the improvements in the directionality reconstruction in the absence of scintillation light we apply the technique to a sample of Cherenkov PEs filtered through a uniform QE of 30\% independent of a wavelength and the default photo-coverage of 65\%. We find that the directionality reconstruction resolution improves from 0.46 to 0.40 radians and from 0.84 to 0.74 radians for the polar angle and azimuthal angles respectively. The inner product improves from 0.78 to 0.83.

\begin{figure}[h!]
    \centering
    \includegraphics[width=0.49\textwidth]{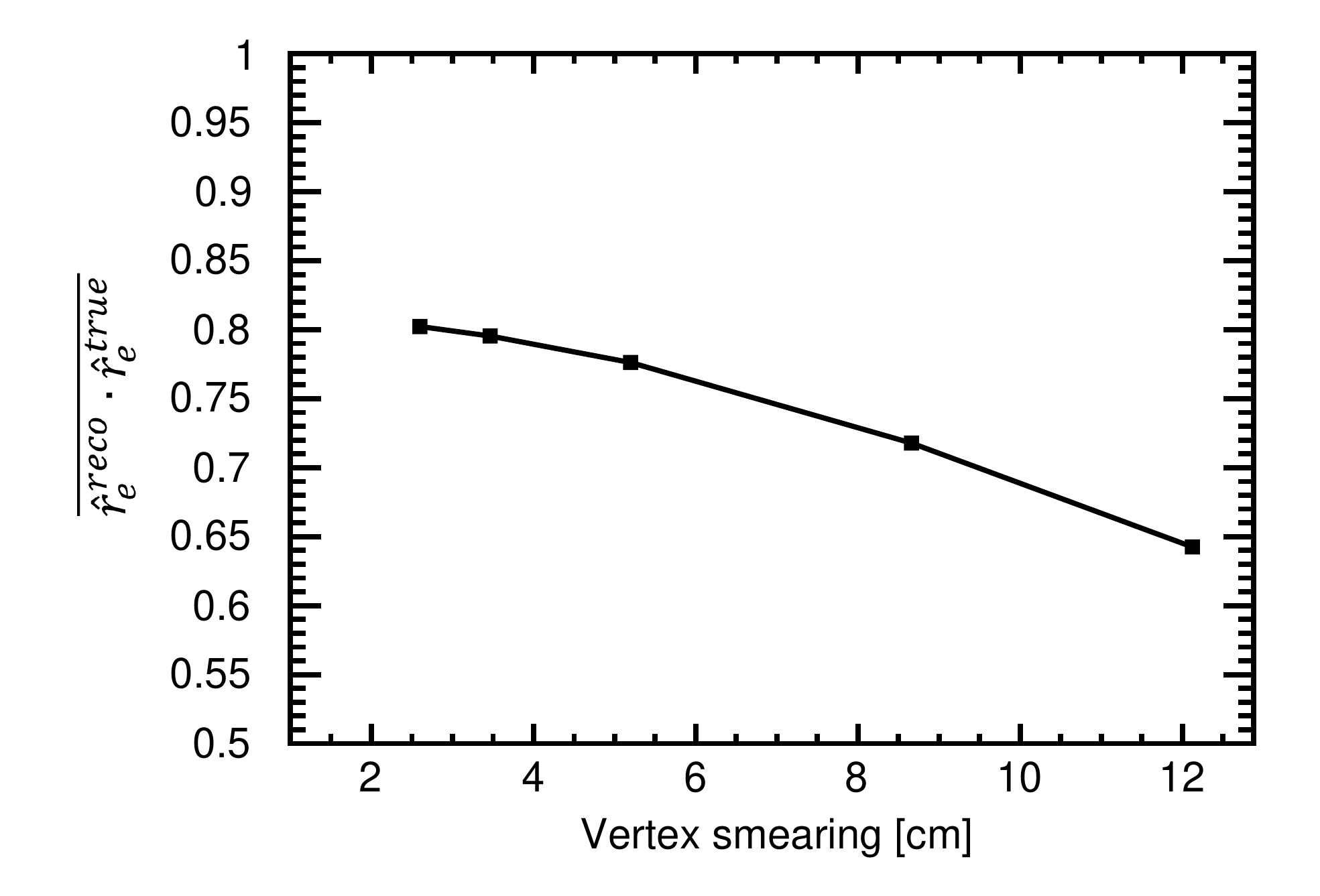}
    \caption{The mean value of the inner product distribution as a function of vertex smearing, $\sigma_{r_v}$.
    The default simulation settings are used except for the vertex smearing, which is varied from 2.6 cm to 12.1 cm.
    (this corresponds to 1.5-7 cm range in $\sigma_{x,y,z}$). 
    }
    \label{fig:direction_smear}
\end{figure}

These improvements are very similar to the case of an ideal light collection. Assuming both photo-coverage and QE of 100\% and keeping all other parameters are the same as in the default simulation, we find the inner product of 0.84. The RMS values of $\Delta \theta = \theta_{reco} - \theta_{true}$ and $\Delta \phi = \phi_{reco} - \phi_{true}$ are 0.39 and 0.72 respectively.

To study the dependence of the directionality reconstruction on the vertex resolution, we run the default simulation at different vertex smearing. Figure~\ref{fig:direction_smear} shows the mean value of the inner product between the reconstructed electron direction and the true electron direction as a function of vertex smearing. The directionality reconstruction degrades as the vertex smearing increases.

Table~\ref{tab:results} summarizes the performance of topological and directional reconstruction for different detector simulation settings.


\begin{table}[h!]
\begin{center}
\begin{tabular}{|p{3.1cm}|p{1.3cm}|p{0.7cm}|p{0.7cm}||p{1cm}|p{1cm}|p{1cm}||p{1.5cm}|p{0.8cm}|p{0.8cm}| }
 \hline
 Simulation scenario & Photo-coverage [\%] & $QE_{che}$ [\%] & $QE_{sci}$ [\%] & 
 AUC & $\frac{1}{FPR@90\%}$ & $\frac{1}{FPR@70\%}$ & $\overline{\hat{r}_e^{reco}\cdot\hat{r}_e^{true}}$ & $RMS_\theta$ & $RMS_\phi$\\
 \hline
 Default simulation& 65 & 12 & 23 & 0.69 & 1.34 & 2.30 & 0.78 & 0.46 & 0.84\\
 Ideal light collection& 100 & 100 & 100 & 0.78 & 1.56 & 3.57 & 0.84 & 0.39 & 0.72\\
 Cherenkov PEs only& 65 & 30 & 0 & 0.77 & 1.59 & 3.39 & 0.83 & 0.40 & 0.74\\
 Same as in Ref.~\cite{harmonics2017}& 100 & 12 & 23 & 0.71 & 1.38 & 2.44 & 0.79 & 0.45 & 0.81\\
 \hline
 \end{tabular}
 \caption{Results of topological and directional reconstruction for different simulation settings. Only photo-coverage and QE deviate from the default simulation; $QE_{che}$ and $QE_{sci}$ are averaged QE for Cherenkov and scintillation light respectively. For the topological reconstruction we list the area under the curve as well as background suppression factors, $\frac{1}{FPR}$, at 90\% and 70\% signal efficiency. For the directionality reconstruction we list the value of the inner product and the value of RMS of the distributions of the difference between the reconstructed and the true polar and azimuthal angles.}
\label{tab:results}
\end{center}
\end{table}

\section{Conclusions}
\label{sec:conclusions}
We have presented a technique to separate \vbb-decay events from \B solar neutrino background in a large liquid scintillator detector. We introduce the Cherenkov-scintillation space-time boundary to increase the contribution from the Cherenkov PEs in the reconstruction of the kinematics of candidate events. The technique has two components. The first component of the technique is the reconstruction of the event topology that
allows the separation of the two-track event topology of \vbb-decay signal events from one-track \B background events. The second component is the reconstruction of the directionality of the one-electron candidate events that
allows for the further suppression of $^8$B background due to the correlation between the direction of scattered electron in $^8$B events and the position of the sun. The directionality reconstruction should be applied after the event topology reconstruction which determines that the topology of a candidate event 
is consistent with the one-track event topology.

We evaluated the performance of the topological and directionality reconstruction separately. For the default detector model with 65\% photo-coverage and assuming vertex resolution of $\sigma_{r_v} = $ 5.2 cm, the method of reconstructing the event topology predicts factors of 1.3 and 2.3 in background suppression at 90\% and 70\% signal efficiency respectively based solely on the reconstruction of the event topology. For the reconstruction of electron directionality we estimate the polar angular resolution to be 0.49 radians and the azimuthal angular resolution to be 0.84 radians. The determination of a combined effect on the sensitivity to the \vbb-decay half-time is a subject of further studies using a detailed detector-specific background model.

For a photo-cathode spectral response modeled after KamLAND PMTs, an increase in photo-coverage beyond 65\% does not lead to significant improvements in the performance of the topological and directionality reconstruction. At the same time an increase in the collection efficiency of Cherenkov PEs relative to scintillation PEs leads to about 50\% higher background suppression factor at 70\% signal efficiency for the topological reconstruction.

Separation of Cherenkov PEs from scintillation PEs is essential for topological and directionality reconstruction in a liquid scintillator detector. The introduced Cherenkov-scintillation space-time boundary provides a mechanism for selecting a sample of the early emitted PEs containing a high fraction of the Cherenkov PEs. The effectiveness of the Cherenkov-scintillation space-time boundary in selecting a PE sample with high fraction of Cherenkov PEs depends on properties of liquid scintillator and performance of light collection system.

Performance of the technique presented in this paper can be improved by developing new liquid scintillators and new light collection systems. Liquid scintillators need to have a narrower emission spectrum shifted to shorter wavelengths and a longer rise time. Light collection systems should include large-area photo-detectors with mm space and 100-psec time resolutions, and red-sensitive photo-cathode. Spectral sorting of photons can minimize the effect of chromatic dispersion and further improve the performance of the technique.

\section{Acknowledgements}
We acknowledge contribution by Lindley Winslow to the earlier version of the algorithm separating \vbb-decay and \B events, and 
AE thanks her for many productive discussions on the event topology reconstruction. We thank Roy Garcia for performing independent cross-checks of calculations related to the spherical harmonics analysis. We thank Henry Frisch for reading this manuscript and providing comments. We thank Evan Angelico, Eric Spieglan, and Chuck Whitmer for feedback and comments. We thank Alexander Bogatskii for discussions on rotation invariants of the SO(3) group. We thank Jo$\Tilde{a}$o Shida for proof reading. AE thanks Ed Blucher, Andy Mastbaum, and Tony LaTorre for discussions on properties of liquid scintillators.
This work is supported by U. S. Department of Energy, Office of Science, Offices of High Energy Physics and Nuclear Physics under contracts DE-SC0008172 and DE-SC0015367; the National Science Foundation under grant PHY-1066014; and the Physical Sciences Division of the University of Chicago. This research was done using resources provided by the Open Science Grid~\cite{Grid1, Grid2}, which is supported by the National Science Foundation award 1148698, and the U.S. Department of Energy's Office of Science. 

\newpage
\bibliography{bibliography.bib}

\newpage
\appendix
\section{Detector Coverage Scheme}
\label{app:coverage}
To uniformly distribute photo-detectors over a sphere, we assign a photo-detector at each vertex of a platonic solid (i.e., tetrahedron, cube, octahedron, dodecahedron and icosahedron). The assigned photo-detectors will be uniformly distributed over the circumscribed sphere of the platonic solid. However, there is no exact solution for distributing more than 20 points over a sphere.

To approximate a uniform distribution of more than 20 photo-detectors over the sphere, we assign  additional photo-detectors to the edges and the faces of the platonic solids and project the photo-detectors onto the circumscribed sphere.

We start with a icosahedron for its largest number of faces among the platonic solids. We equally divide each edge by $N$ times and consequently divide each triangular face by $N^2$ times.

\begin{figure*}
\centering
\label{fig:coverage_1}
\includegraphics[width=0.49\textwidth]{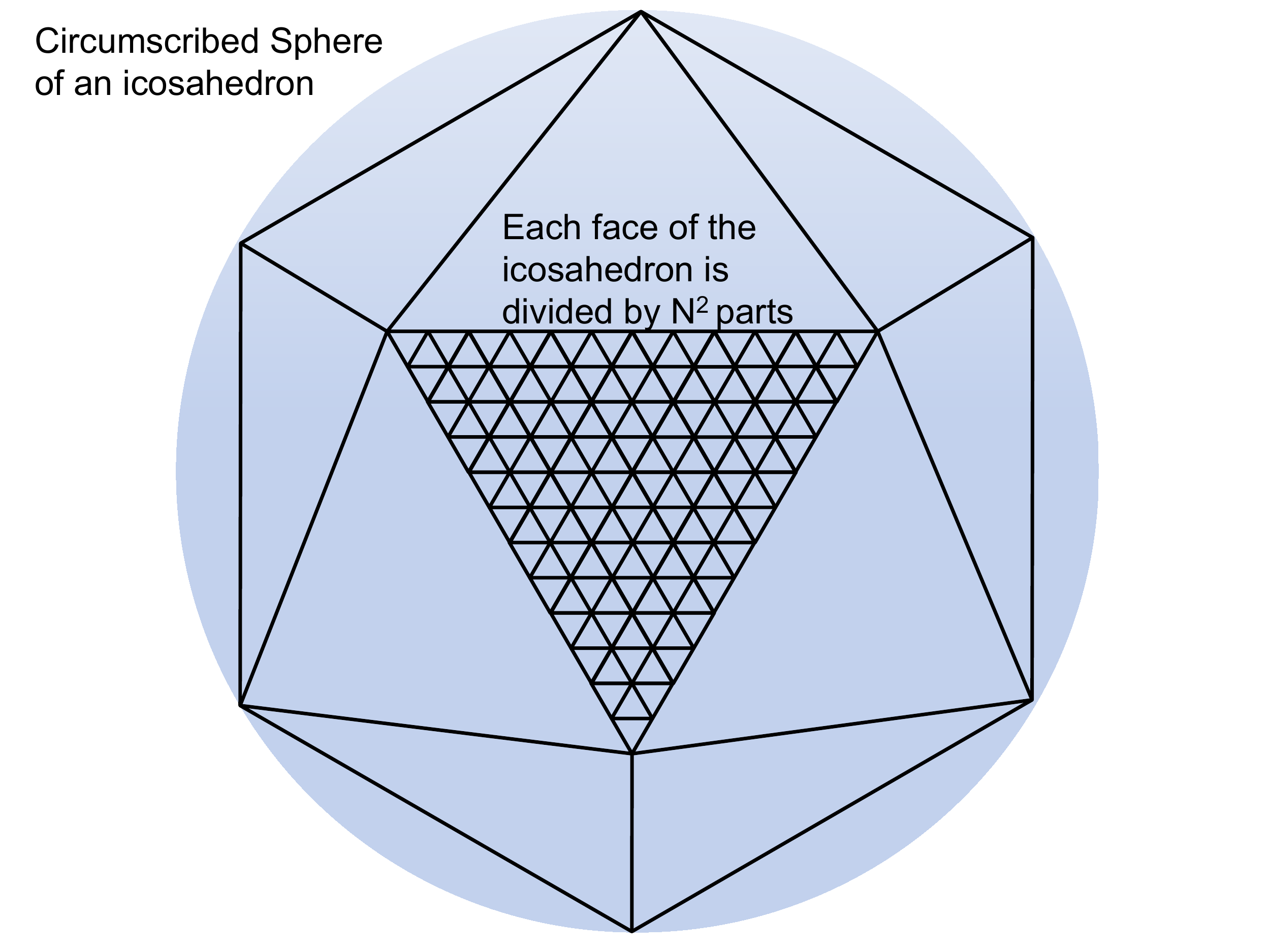}%
\label{fig:coverage_2}
\includegraphics[width=0.49\textwidth]{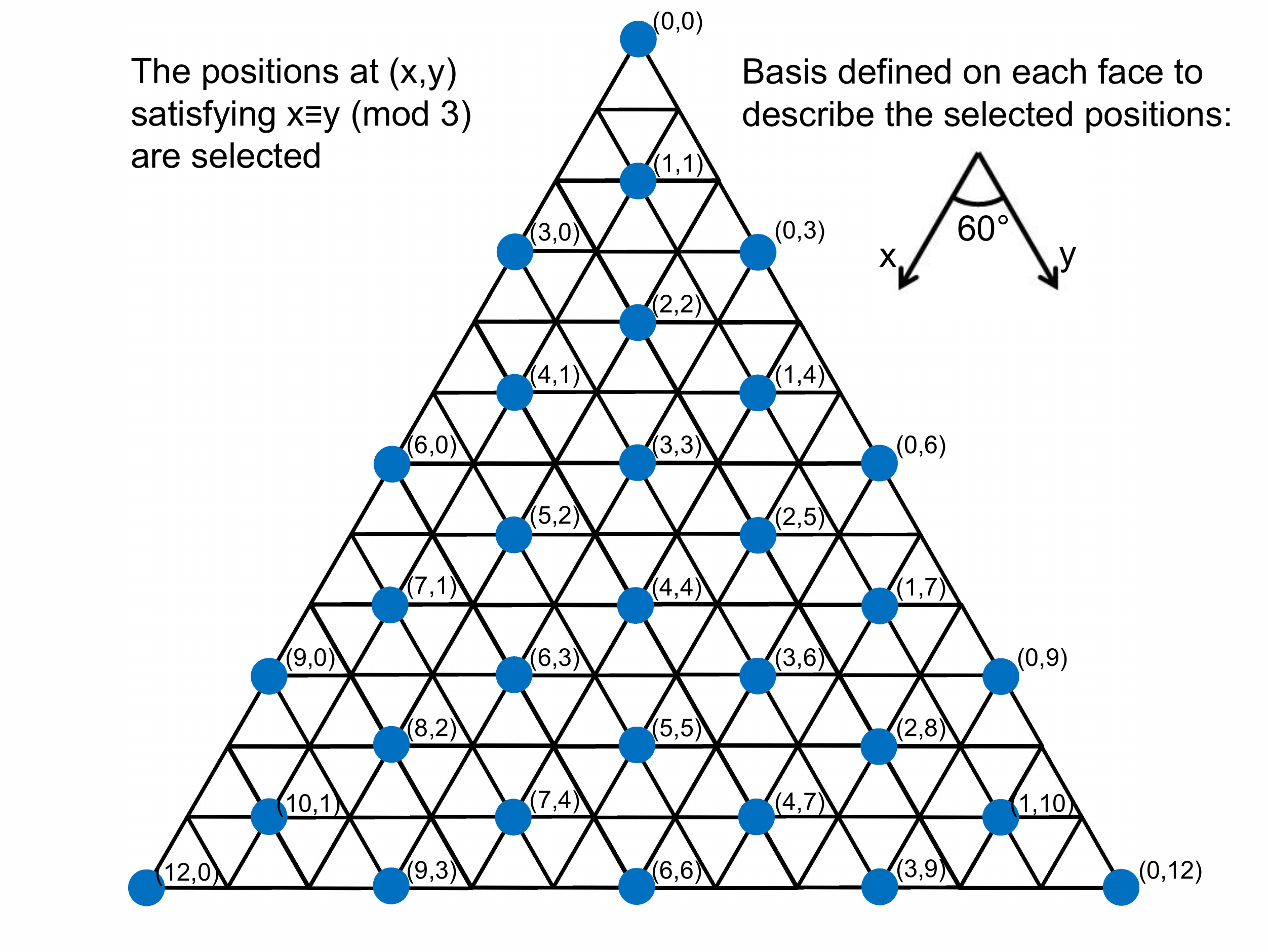}%
\vskip\baselineskip
\vskip\baselineskip
\vskip\baselineskip
\label{fig:coverage_3}
\includegraphics[width=0.49\textwidth]{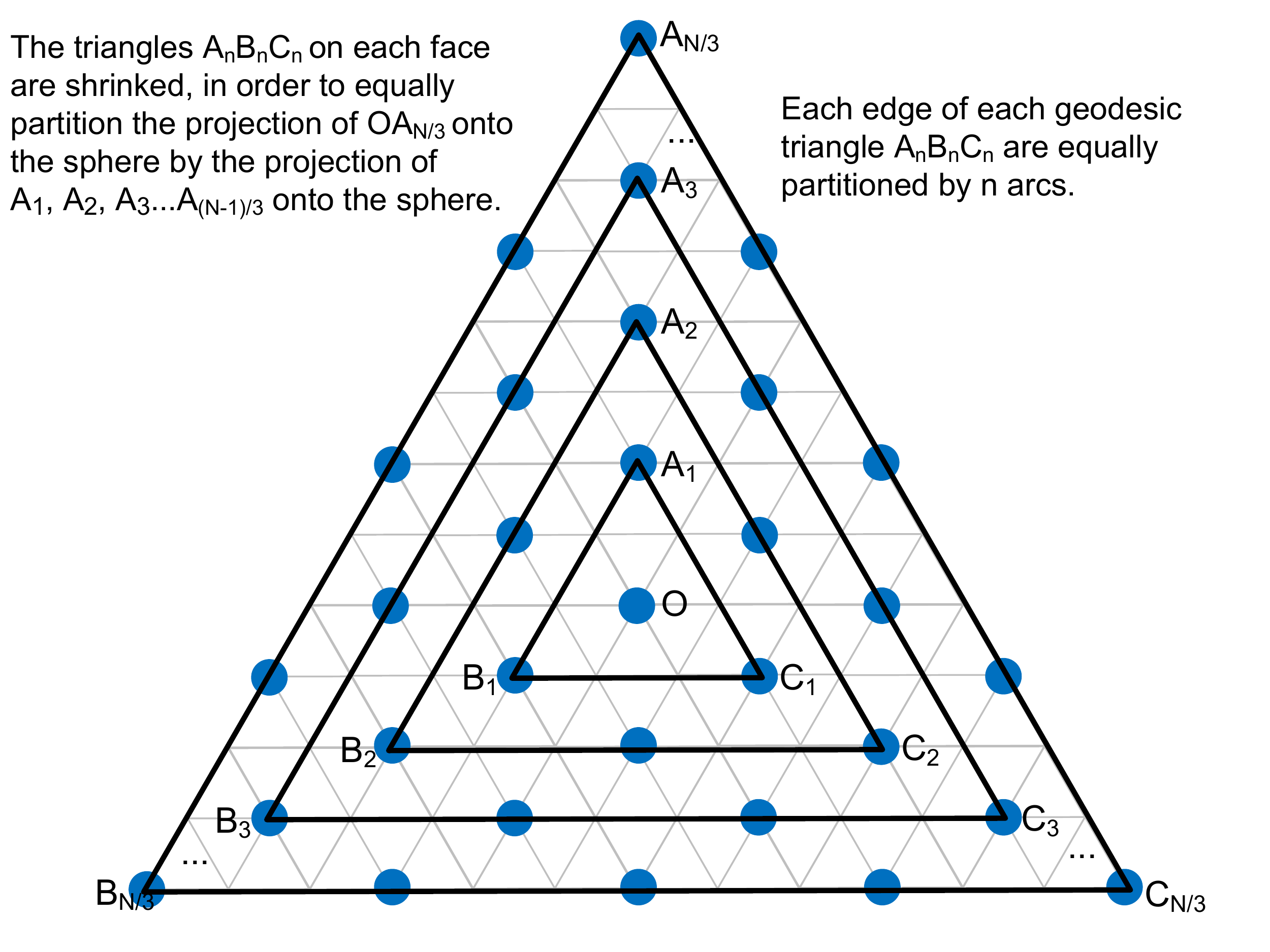}%
\label{fig:coverage_4}
\includegraphics[width=0.49\textwidth]{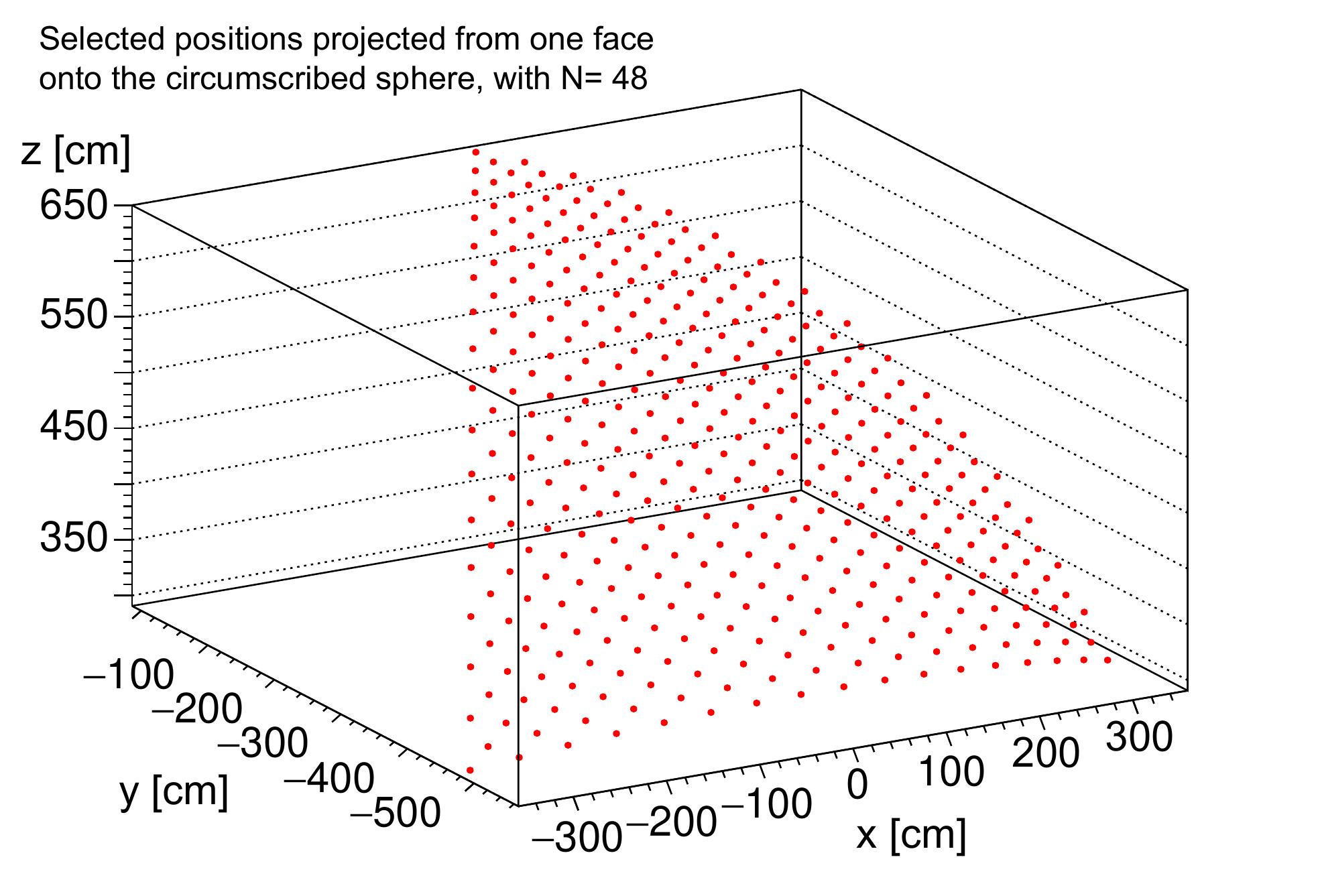}%
\centering
\caption{[\emph{Top left}]: An icosahedron circumscribed by a sphere. We subdivide each face by $N^2$ times (where $N\equiv 0 (mod 3)$) and select some particular positions on each face and project the positions onto the circumscribed sphere to enable the assignment of more detectors.
[\emph{Top Right}]: The selection of the detectors on each face on the icosahedron before the projection. We assign detectors to all the positions with coordinates $(x,y)$ satisfying $x\equiv y (mod 3)$.
[\emph{Bottom Left}]: An illusration of tuning the positions of the detectors. The triangles are rescaled such that the vertices of the triangles are equally spaced over the sphere. Furthermore, the geodesic projected from $A_n B_n$ $B_n C_n$ $C_n A_n$ are respectively equally partitioned by $n$ times.
[\emph{Bottom Right}]: The positions of the detectors over the sphere projected from one face of the icosahedron, given the subdivision number $N=48$.
}
\label{fig:coverage_scheme}
\end{figure*}

We introduce the following coordinate system. Starting from one vertex of a face, we define the directions along the two edges to be unit vectors $\vec{e}_1$ and $\vec{e}_2$. We also define the length of each division to be 1. Consequently, all the intersecting points in the Figure~\ref{fig:coverage_scheme} have unique integer coordinates $(x,y)$ under the basis of $\vec{e}_1$ and $\vec{e}_2$. We select all the points that satisfy $x\equiv y \pmod{3}$.

However, for such selection, the distance between two closest projected points near the vertex is smaller than the distance between two closest projected points near the center of the face (see Figure~\ref{fig:coverage_scheme}).

Therefore, in order to space the points more uniformly, we first partition all the points by classes of triangles (see Figure~\ref{fig:coverage_scheme}, the triangles with the thickened line width are the subdivided triangle classes). Notice that any straight lines in $\mathbb{R}^3$ projected onto a sphere $S^2$ are geodesics; therefore, the projected triangle classes onto the sphere are geodesic triangles, regardless of the size of the triangle.

While fixing the common centroid of these triangle classes unchanged, we scale all the triangle classes such that the vertices of these triangle classes are equally spaced along the geodesics that connect the icosahedron vertex and the common centroid. Meanwhile, given the total number of points along each edge of the subdivided triangle class remains unchanged, we reassign the location of the points along each edge of each subdivided triangle, such that the points are equally spaced along the edge of each triangle class after being projected onto the circumscribed sphere (see Figure~\ref{fig:coverage_scheme}).

\section{Spherical Harmonics Analysis}
\label{app:harmonics}
\subsection{S-spectrum $S_{\ell}$}
Suppose an angular function $f(\theta,\phi)$ on a unit sphere $\mathbb{S}^{2}$ represents the angular distribution of photo-electrons (PE) over the detector surface. Such angular function $f(\theta,\phi)$ has a corresponding spherical harmonics expansion:
\begin{eqnarray}
\label{EqB:detection_signal}
f(\theta,\phi) = \sum_{\ell=0}^{+\infty} \sum_{m=-\ell}^{\ell} f_{\ell m} Y_{\ell m}(\theta,\phi)
\end{eqnarray}

where $Y_{\ell m}(\theta,\phi)$ are tesseral harmonics (real-valued spherical harmonics), which form a complete orthonormal basis of the rigged Hilbert space
(all real square integrable functions and Dirac delta functions defined over $S^2$):
\begin{eqnarray}
\label{EqB:spherical_harmonics}
Y_{\ell m}(\theta,\phi) = \left\{
  \begin{array}{@{}ll@{}}
    \sqrt{2}\sqrt{\frac{(2\ell+1)}{4\pi} \frac{(\ell-m)!}{(\ell+m)!}}P_{\ell}^m(\cos\theta)\cos~m\phi &, \text{if}\ m>0 \\
    \sqrt{\frac{(2\ell+1)}{4\pi} \frac{(\ell-m)!}{(\ell+m)!}}P_{\ell}^m(\cos\theta) &, \text{if}\ m=0 \\
    \sqrt{2}\sqrt{\frac{(2\ell+1)}{4\pi} \frac{(\ell-m)!}{(\ell+m)!}}P_{\ell}^{|m|}(\cos\theta)\sin~|m|\phi &, \text{if}\ m<0
  \end{array}\right.
\end{eqnarray}

and the projection coefficients $f_{\ell m}$ are given by the projection of $f(\theta,\phi)$ onto the basis $Y_{\ell m}(\theta,\phi)$:
\begin{eqnarray}
\label{EqB:expansion_coefficient}
f_{\ell m}=\int_{0}^{\pi}\sin \theta d\theta\int_{0}^{2\pi}d\phi f(\theta,\phi)Y_{\ell m}(\theta, \phi)
\end{eqnarray}

For a given angular functuion $f(\theta,\phi)$, its rotationally invariant S-spectrum $S_{\ell}$ is defined as:
\begin{eqnarray}
\label{EqB:s_spectrum}
S_{\ell} = \sum_{m=-\ell}^{\ell} |f_{\ell m}|^2
\end{eqnarray}

\subsection{Weighted Angular Distribution $f(\theta,\phi)$ of Photo-electrons}
Suppose the Dirac delta function $\delta(\cos\theta-\cos\theta_{i})\delta(\phi-\phi_{i})$ represents the angular distribution of an individual PE at a location with coordinates $(\theta_{i},\phi_{i})$ on the detector surface. The $L^{1}$-normalized (see Equation~\ref{EqB:normalization_L1}) weighted angular distribution $f(\theta,\phi)$ of all $N_{PE}$ photo-electrons (PEs) is
\begin{eqnarray}
\label{EqB:normalized_detection_signal}
f(\theta,\phi) =\frac{ \sum_{i=1}^{N_{PE}}W(t_i,\theta_i)\delta(\cos\theta-\cos\theta_{i})\delta(\phi-\phi_{i})}{\sum_{i=1}^{N_{PE}}W(t_i,\theta_i)}
\end{eqnarray}

Substituting the distribution $f(\theta,\phi)$ into the Equation~\ref{EqB:expansion_coefficient} yields the corresponding projection coefficients $f_{\ell m}$:
\begin{eqnarray}
\label{EqB:simplified_coefficient}
f_{\ell m}=
\int_{0}^{\pi} \sin\theta d\theta\int_{0}^{2\pi}d\phi \frac{\sum_{i=1}^{N_{PE}}W(t_i,\theta_i)\delta(\cos\theta-\cos\theta_{i})\delta(\phi-\phi_{i})}
{\sum_{i=1}^{N_{PE}}W(t_i,\theta_i)}
Y_{\ell m}(\theta, \phi)
=\frac{\sum_{i=1}^{N_{PE}}W(t_i,\theta_i) Y_{\ell m}(\theta_{i}, \phi_{i})}{\sum_{i=1}^{N_{PE}}W(t_i,\theta_i)}
\end{eqnarray}

and also its S-spectrum $S_{\ell}$:
\begin{eqnarray}
\label{EqB:simplified_s_spectrum}
S_{\ell} = \sum_{m=-\ell}^{\ell}\abs{f_{\ell m}}^2
=\sum_{m=-\ell}^{\ell}\abs{\frac{\sum_{i=1}^{N_{PE}}W(t_i,\theta_i) Y_{\ell m}(\theta_{i}, \phi_{i})}{\sum_{i=1}^{N_{PE}}W(t_i,\theta_i)}}^2
=\frac{\sum_{m=-\ell}^{\ell} \abs{\sum_{i=1}^{N_{PE}}W(t_i,\theta_i)Y_{\ell m}(\theta_{i}, \phi_{i})}^2}{\abs{\sum_{i=1}^{N_{PE}}W(t_i,\theta_i)}^{2}}
\end{eqnarray}

\subsection{Normalization of the Angular Distribution $f(\theta,\phi)$}
In this paper, the angular distribution of PEs, $f(\theta,\phi)$, is normalized by the $L^{1}$ norm:
\begin{eqnarray}
\label{EqB:normalization_L1}
||f(\theta,\phi||_{1}= \int_{S^{2}}d\Omega \abs{f(\theta,\phi)}
\end{eqnarray}

For example, the $L^{1}$ norm of the weighted angular distribution $f(\theta,\phi)=\sum_{i=1}^{N_{PE}}W(t_i,\theta_i)\delta(\cos\theta-\cos\theta_{i})\delta(\phi-\phi_{i})$, which is not yet normalized, is $\sum_{i=1}^{N_{PE}}W(t_i,\theta_i)$. As a consequence, in this paper, all the S-spectra has the normalization factor $\abs{\sum_{i=1}^{N_{PE}}W(t_i,\theta_i)}^2$ in the denominator.

Note, the sum of $S_{\ell}$'s over all multiple moments $\ell$ equals to the $L^{2}$ norm of the function $f(\theta,\phi)$:
\begin{eqnarray}
\label{EqB:normalization_L2}
||f(\theta,\phi||_{2} = \int_{S^{2}} d\Omega|f(\theta,\phi)|^2 = \sum_{\ell=0}^{+\infty} S_{\ell}
\end{eqnarray}

However, the detection signal $f(\theta,\phi)$ in this paper, which is a weighted sum of the Dirac delta functions, is not square integrable; as a consequence, the sum of $S_{\ell}$'s diverges.

\subsection{Rotation Invariance of the S-spectrum}
To simplify the notation in the following proof, we adopt the Dirac bracket notation: $\ket{f}=f(\theta,\phi)$ and $\ket{Y_{\ell m}}=Y_{\ell m}(\theta,\phi)$.

We define operators $\hat{L}_x$, $\hat{L}_y$, $\hat{L}_z$, $\hat{L}^2$, $\hat{R}$ to be:
\begin{eqnarray}
\label{EqB:L_operator}
\left\{
  \begin{array}{@{}ll@{}}
    \hat{L}_x=y\partial_z-z\partial_y=-\sin\phi\partial_\theta-\cot\theta\cos\phi\partial_\phi\\
    \hat{L}_y=z\partial_x-x\partial_z=\cos\phi\partial_\theta-\cot\theta\sin\phi\partial_\phi\\
    \hat{L}_z=x\partial_y-y\partial_x=\partial_\phi\\
    \hat{L}^2=\frac{\partial_\theta(\sin\theta\partial_\theta)}{\sin\theta}+\frac{\partial^2_\phi}{\sin^2\theta}\\
    \hat{R}=\hat{R}'_z\hat{R}_x\hat{R}_z=e^{L_z \Delta \phi_z'}e^{L_x \Delta \phi_x}e^{L_z \Delta \phi_z}
  \end{array}\right.
\end{eqnarray}

The tesseral harmonics $\ket{Y_{\ell m}}$ (real-valued spherical harmonics) are the eigenvectors of $\hat{L}^2$: $\hat{L}^2\ket{Y_{\ell m}}=\ell(\ell+1)\ket{Y_{\ell m}}$. The collection of the tesseral harmonics forms a complete orthonormal basis of the rigged Hilbert space $\mathscr{H}$ over $\mathbb{S}^2$. We define $\mathscr{H}_{\ell}$ to be a subspace of the space $\mathscr{H}$, which is spanned by the eigenvectors $\ket{Y_{\ell m}}$ with a given fixed $\ell$. The collection of $\ket{Y_{\ell m}}$ with a given fixed $\ell$ forms a complete orthonormal basis of $\mathscr{H}_{\ell}$. Therefore, the operator $\sum_{m=-\ell}^{+\ell} \ket{Y_{\ell m}}\bra{Y_{\ell m}}$ is equivalent to the identity operator $\hat{I}_{\ell}$ in $\mathscr{H}_{\ell}$ (and equivalent to 0 in the other subspaces $\mathscr{H}_{\ell'}$ with $\ell' \neq \ell$).

$[\hat{L}_i,\hat{L}^2]=0$ leads to
$[\hat{R}_i,\hat{L}^2]=0$ and $[\hat{R},\hat{L}^2]=0$. 
Therefore, $\hat{R}\ket{Y_{\ell m}}$ is an eigenstate of $\hat{L}^2$ in $\mathscr{H}_{\ell}$.

Using the Dirac bracket notation, we rewrite the definition of the S-spectrum $S_{\ell}$:
\begin{eqnarray}
\label{EqB:Dirac_S_spectrum}
S_{\ell}=\sum_{m=-\ell}^{+\ell}\abs{f_{\ell m}}^2=\sum_{m=-\ell}^{+\ell}\bra{f}\ket{Y_{\ell m}}\bra{Y_{\ell m}}\ket{f}
\end{eqnarray}

Applying an arbitrary rotation $\hat{R}$ to the detection signal $\ket{f'}=\hat{R}\ket{f}$, the S-spectrum $S_{\ell}$ is transformed to be $S_{\ell}'$:
\begin{eqnarray}
\label{EqB:proof_rotation_invariance_step1}
S_{\ell}'=\sum_{m=-\ell}^{+\ell}\abs{f'_{\ell m}}^2=\sum_{m=-\ell}^{+\ell}\bra{f'}\ket{Y_{\ell m}}\bra{Y_{\ell m}}\ket{f'}=\sum_{m=-\ell}^{+\ell}\bra{f}\hat{R}^{\dagger}\ket{Y_{\ell m}}\bra{Y_{\ell m}}\hat{R}\ket{f}
\end{eqnarray}

Applying spectral decomposition to $\ket{f}$ and organizing:
\begin{eqnarray}
\label{EqB:proof_rotation_invariance_step2}
S_{\ell}'=\sum_{m=-\ell}^{+\ell} \sum_{\ell' m'} \sum_{\ell'' m''} f_{\ell' m'} f_{\ell'' m''} \bra{Y_{\ell' m'}}\hat{R}^{\dagger}\ket{Y_{\ell m}}\bra{Y_{\ell m}}\hat{R}\ket{Y_{\ell'' m''}}
\end{eqnarray}

Recall $\hat{R}\ket{Y_{\ell m}}$ is an eigenstate of $\hat{L}^2$ in $\mathscr{H}_{\ell}$. Therefore, $\bra{Y_{\ell m}}\hat{R}\ket{Y_{\ell'' m''}}$ is non-zero, if $\ell''=\ell$. Similarly, $\bra{Y_{\ell m}}\hat{R}\ket{Y_{\ell' m'}}$ is non-zero, if $\ell'=\ell$. Consequently, the only non-zero term in the sum above is
\begin{eqnarray}
\label{EqB:proof_rotation_invariance_step3}
S_{\ell}'=\sum_{m',m''} f_{\ell m'}f_{\ell m''}\bra{Y_{\ell m'}}\hat{R}^{\dagger}\sum_{m=-\ell}^{+\ell}\ket{Y_{\ell m}}\bra{Y_{\ell m}}\hat{R}\ket{Y_{\ell m''}}
\end{eqnarray}

Recall $\sum_{m=-\ell}^{+\ell}\ket{Y_{\ell m}}\bra{Y_{\ell m}}$ is equivalent to the identity operator, when applying to any vectors in $\mathscr{H}_{\ell}$. Therefore, the equation above can be simplified:
\begin{eqnarray}
\label{EqB:proof_rotation_invariance_step4}
S_{\ell}'=\sum_{m',m''} f_{\ell m'}f_{\ell m''}\bra{Y_{\ell m'}}\hat{R}^{\dagger}\hat{I}_{\ell}\hat{R}\ket{Y_{\ell m''}}=\sum_{m',m''} f_{\ell m'}f_{\ell m''}\bra{Y_{\ell m'}}\ket{Y_{\ell m''}}=\sum_{m',m''} f_{\ell m'}f_{\ell m''}\delta_{m' m''}=\sum_{m'} f_{\ell m'}f_{\ell m'}
\end{eqnarray}

Changing the label $m'$ to $m$, we prove the rotation invariance of the S-spectrum $S_{\ell}$:
\begin{eqnarray}
\label{EqB:proof_rotation_invariance_step5}
S_{\ell}'=\sum_{m} f_{\ell m}f_{\ell m}=\sum_{m} \abs{f_{\ell m}}^2=S_{\ell}
\end{eqnarray}

\section{Maximum Likelihood Estimation}
\label{app:likelihood}
\subsection{Conditional Probability Distribution}
For an arbitrary event, given its measurement $\vec{S}_{meas}=(S_1,\alpha)$ in the parameter space of $(S_1,\alpha)$, the likelihood for the event to be a signal (0\nbb, $2e$) is the conditional probability $p(1e|\vec{S}_{meas})$. Similarly, the likelihood for the event to be a background (solar neutrino interaction, $1e$) is the conditional probability $p(2e|\vec{S}_{meas})$. Based on Baysian theorem, these posterior probabilities are given by:
\begin{eqnarray}
\label{eqC:Baysian}
\left\{
  \begin{array}{@{}ll@{}}
    p(1e|\vec{S}_{meas})&=\frac{p(1e)}{p(\vec{S}_{meas})}p(\vec{S}_{meas}|1e)=k_{1}p(\vec{S}_{meas}|1e)\\
    p(2e|\vec{S}_{meas})&=\frac{p(2e)}{p(\vec{S}_{meas})}p(\vec{S}_{meas}|2e)=k_{2}p(\vec{S}_{meas}|2e)
  \end{array}\right.
\end{eqnarray}
where the response function $p(\vec{S}_{meas}|1e)$ ($p(\vec{S}_{meas}|2e)$) is the conditional probability of getting measurement $\vec{S}_{meas}$ given the event is a 0\nbb~signal (\B background), which can be read from the histograms in the Figure~\ref{fig:Parameter_Space}. The coefficients $k_1$ and $k_2$ are determined by the ratio of the prior probabilities $p(1e):p(2e)$ and the normalization. In this paper, the ratio of the prior probabilities ${k_1}:{k_2}={p(1e)}:{p(2e)}$ is set to be 1.
\begin{eqnarray}
\label{eqC:prior}
\left\{
  \begin{array}{@{}ll@{}}
    {k_1}:{k_2}={p(1e)}:{p(2e)}=1\\
	k_{1}p(\vec{S}_{meas}|1e)+k_{2}p(\vec{S}_{meas}|2e)=1
  \end{array}\right.
\end{eqnarray}

Therefore, through solving the four equations in the Equation~\ref{eqC:Baysian} and~\ref{eqC:prior}, the posterior probabilities $p(1e|\vec{S}_{meas})$ and $p(2e|\vec{S}_{meas})$ are given by
\begin{eqnarray}
\label{eqC3}
\left\{
  \begin{array}{@{}ll@{}}
    p(1e|\vec{S}_{meas})&=\frac{p(\vec{S}_{meas}|1e)}{p(\vec{S}_{meas}|1e)+p(\vec{S}_{meas}|2e)}\\
    p(2e|\vec{S}_{meas})&=\frac{p(\vec{S}_{meas}|2e)}{p(\vec{S}_{meas}|1e)+p(\vec{S}_{meas}|2e)}
  \end{array}\right.
\end{eqnarray}

\subsection{Multi-dimensional Probability Distribution}
In practice, filling the histograms of $p(s_1,\alpha|1e)$ and $p(s_1,\alpha|2e)$ requires more data than filling the histograms of $p(s_1|1e)$, $p(\alpha|1e)$, $p(s_1|2e)$ and $p(\alpha|2e)$. Given limited data, 
we 
replace these multi-dimensional histograms by the product of multiple 1-dimensional histograms:
\begin{eqnarray}
\label{eqC:approximate_likelihood}
\left\{
  \begin{array}{@{}ll@{}}
	p(S_1,\alpha|1e)\approx p(S_1|1e)p(\alpha|1e)\\
	p(S_1,\alpha|2e)\approx p(S_1|2e)p(\alpha|2e)
  \end{array}\right.
\end{eqnarray}

Based on the approximation above, the probabilities in the Equation~\ref{eqC3} become:
\begin{eqnarray}
\label{eqC:off_center_likelihood}
    p(2e|\vec{S}_{meas})=\frac{p(\vec{S}_{meas}|2e)}{p(\vec{S}_{meas}|1e)+p(\vec{S}_{meas}|2e)}=\frac{p(S_1|2e)p(\alpha|2e)}{\sum_{j=1}^{2}p(S_1|je)p(\alpha|je)}
\end{eqnarray}


\end{document}